\documentclass{emulateapj}

\newcommand{\sci}[1]{\ensuremath{\times 10^{#1}}}
\newcommand{\rsun}{R$_{\odot}$}

\slugcomment{Accepted to \apj}
\shorttitle{The $\beta$ Leo Debris Disk}
\shortauthors{Stock et al.}

\begin{document}

\title{The structure of the $\beta$ Leonis Debris Disk}
\author{Nathan D. Stock\altaffilmark{1}, Kate Y.L. Su\altaffilmark{1},
  Wilson Liu\altaffilmark{2}, Phil M. Hinz\altaffilmark{1}, George
  H. Rieke\altaffilmark{1}, Massimo Marengo\altaffilmark{3}, Karl
  R. Stapelfeldt\altaffilmark{4}, Dean C. Hines\altaffilmark{5}, and
  David E. Trilling\altaffilmark{6}}  

\altaffiltext{1}{Steward Observatory, University of Arizona, 933 N
  Cherry Avenue, Tucson, AZ 85721, USA; nstock@as.arizona.edu}
\altaffiltext{2}{Infrared Processing and Analysis Center, California 
  Institute of Technology, Mail Code 100-22, Pasadena, CA 91125, USA}
\altaffiltext{3}{Department of Physics and Astronomy, Iowa State
  University, Ames, IA 50010, USA}
\altaffiltext{4}{JPL/Caltech, 4800 Oak Grove Drive, Pasadena, CA
  91109}
\altaffiltext{5}{Space Science Institute, 4750 Walnut Street, Suite
  205 Boulder, CO 80301, USA} 
\altaffiltext{6}{Department of Physics and Astronomy, Northern Arizona
University, 602 S Humphreys Street, Flagstaff, AZ 86011, USA}

\begin{abstract}

We combine nulling interferometry at 10 $\mu$m using the MMT and Keck
Telescopes with spectroscopy, imaging, and photometry from 3 to 100
$\mu$m using {\it Spitzer} to study the debris disk around $\beta$ Leo
over a broad range of spatial scales, corresponding to radii of 0.1 to
$\sim$100 AU. We have also measured the close binary star {\it o} Leo
with both Keck and MMT interferometers to verify our procedures with
these instruments. The $\beta$ Leo debris system has a complex
structure: 1.) relatively little material within 1 AU; 2.) an inner
component with a color temperature of $\sim$600 K, fitted by a dusty
ring from about 2 to 3 AU; and 3.) a second component with a color
temperature of $\sim$120 K fitted by a broad dusty emission zone
extending from about $\sim$5 AU to $\sim$55 AU. Unlike many other
A-type stars with debris disks, $\beta$ Leo lacks a dominant outer
belt near 100 AU.

\end{abstract} 

\keywords{circumstellar matter -- infrared: stars -- planetary systems -- 
stars: individual ($\beta$ Leo, {\it o} Leo) -- techniques: interferometric}

\section{Introduction}

The {\it IRAS} discovery of excess infrared radiation from many stars
introduced a powerful new approach to study neighboring planetary
systems. The dust that radiates in the infrared in these "planetary
debris systems" is cleared away on timescales much shorter than the
lifetime of the host stars, through processes such as radiation
pressure and Poynting-Robertson drag. Its presence indicates the
existence of larger bodies that replenish it through cascades of
collisions \citep{BP93}. These bodies may be organized into large
circumstellar structures analogous to the asteroid and Kuiper Belts in
the Solar System. Like them, this extrasolar debris may be shepherded
by the gravity of large planets. Moreover, the dust location is a
tracer of the non-gravitational drag and ejection forces, as well as
of other effects such as sublimation if the dust approaches the star
too closely. The structure of a relatively easily detected planetary
debris system therefore reveals clues about the nature of the
undetectable planetary system that sustains the debris.

Although readily detected with cryogenic space infrared telescopes
(e.g., {\it IRAS, ISO, Spitzer,} and {\it Herschel}), only a small fraction
of debris systems have well-understood structures. The small apertures
and long operating wavelengths of these infrared telescopes provide
inadequate resolution to resolve the majority of known debris systems
well. Scattered light imaging, particularly with {\it HST}, can only
reach disks of high optical depth \citep{K05,wyatt08}. Ground-based
imaging in the infrared can provide geometric information such as
width, inclination and the presence of rings, warps or asymmetries
\citep[e.g.,][]{telesco05}, but is sensitive only to
high-surface-brightness structures. Spectral data can provide
information on the chemical composition and temperature of a debris
disk; however, it leaves the grain size and distance from the central
star poorly determined, since both are derived from the grain optical
properties. Combining infrared spectral and imaging data is necessary
to decode the information that debris disks can provide about
planetary systems.

Different spectral regions preferentially probe different radial zones
around the central star. Simple arguments from the Wien Displacement
Law and radiative equilibrium show, for example, that 2 $\mu$m
measurements should be dominated by dust at $\sim$1500 K, which would
lie at several tenths of AUs from a luminous A-type star. Similarly,
observations at 10 $\mu$m will preferentially be sensitive to dust at
a distance of 2--5 AU, those at 24 $\mu$m will be dominated by
structures at 10--20 AU, and measurements at 70 $\mu$m are best suited
to probe in the $\sim$100 AU region, which for most debris disks
appears to be the outermost zone. A consistent combination of
observations over the entire spectral range of 2 to 100 $\mu$m is
required to build a reasonably comprehensive model of a debris
system. Only a few systems have been modeled in this way, such as
Vega \citep{su05,absil06}.

In this paper, we report observations of $\beta$ and {\it o} Leo with
the Keck Interferometer Nuller (KIN), the Bracewell Infrared Nulling
Cryostat (BLINC) on the MMT, and with {\it Spitzer}. Both KIN and
BLINC are operated at 10 \micron\ and trace material close to the
star: KIN has a resolution of 0\farcs012 and field of view of
$\sim$0\farcs5, which is complemented by BLINC observations at a resolution
of 0\farcs2 and a field of view of 1\farcs5. {\it Spitzer} has a beam
of 6\arcsec~at 24 \micron\ and 18\arcsec~at 70 \micron. Together these
three facilities can search for extended dust emission 
from 0\farcs01 to $\sim$10\arcsec, corresponding to 0.1--100 AU.

We present the observations in Section \ref{obs} and general results
in Section \ref{results}. In Section \ref{analyses} we discuss and
analyze all available data of $\beta$ Leo to set constraints on the
circumstellar material. We then build a physical disk model based on
these derived properties with assumed grain properties in Section
\ref{physical_model}, followed by discussion of the results and a
conclusion in Section \ref{discussion} and Section \ref{conclusion}.

\section{Observations And Data Reductions} 
\label{obs}

\subsection{BLINC}

Details about the method of nulling interferometry employed by BLINC
can be found in the appendix. Briefly, the method involves delaying a
beam from one aperture by half a wavelength and then overlapping it
with a beam from a second aperture. The destructive interference that
results can be arranged to suppress the light coming from the central
star, allowing faint, extended emission to be observed. Specifically, 
BLINC is sensitive between radii of $\sim$0\farcs12 and 0\farcs8. 

Observations were carried out on 2008 March 27 at the 6.5 m MMT,
located on Mount Hopkins, AZ. They consisted of taking ten 1-s
frames on targets and calibrators with BLINC set to 
interfere the apertures destructively. These measurements were followed 
by nodding the telescope 15\arcsec ~to take ten 1-s background
frames. The process was then repeated with BLINC set to 
interfere the apertures constructively. To maximize observing efficiency,
40-frame sets of destructive images were occasionally taken for
$\beta$ and {\it o} Leo, since these frames had low signal to noise,
while constructive images were skipped. All of the observations reported 
here used an N-band filter, covering roughly 8 to 13 \micron. 

Table \ref{tab-mmtobs} lists the number of frames taken for each star
in both destructive and constructive modes of BLINC, as well as the
number of usable frames. 
We discarded the observed frames when the adaptive optics
(AO) system broke lock in the middle of taking a dataset (often due to
wind gusts), and when BLINC did not properly set the pathlength
difference between beams. In addition,
some frames taken at the end of observing $\beta$ and {\it o} Leo, as
well as all the frames taken on the calibrator star $\alpha$ Boo were
adversely affected by weather conditions and were discarded. The
results of comparing the weather-affected $\beta$ and {\it o} Leo data
with the $\alpha$ Boo data are consistent with the data presented here,
but with substantially larger error bars.  

\begin{deluxetable*}{ccccc}
\tablewidth{0pt}
\tablecaption{Summary of Observations at MMT\tablenotemark{a} \label{tab-mmtobs}
}
\tablehead{
\colhead{Star} & \colhead{Total frames} & \colhead{Total
    frames} & \colhead{Used frames} & \colhead{Used frames} \\   
\colhead{} & \colhead{destructive} & \colhead{constructive} &
\colhead{destructive} & \colhead{constructive}
}  
\startdata
$\beta$ Gem & 160 & 150 & 79 & 80 \\
$\beta$ Leo & 480 & 200 & 260 & 70 \\
{\it o} Leo & 720 & 150 & 680 & 120 \\
$\alpha$ Boo\tablenotemark{b} & 50 &50 & 0 & 0 \\
\enddata 
\tablenotetext{a}{Data taken at the MMT on March 27, 2008. For
  each set of on-star observations, an equal number of sky background
  frames were taken.}  
\tablenotetext{b}{$\alpha$ Boo data was not used due to weather
  problems, see text.}  
\end{deluxetable*} 

\subsection{KIN}

Observations of $\beta$ Leo and {\it o} Leo were obtained with the
Keck Interferometer Nuller (KIN) during runs on 2008 February 16--18
and 2008 April 14--16.  The 85 m baseline on KIN provides the ability
to detect circumstellar material in the N-band at angular separations
of a few milliarcseconds from the primary star, within the KIN field
of view defined as an elliptical Gaussian with FWHM of
0\farcs5x0\farcs44 \citep{colavita09}. Thus, the observations 
taken by KIN are spatially complementary to the shorter baseline of
BLINC. Each primary aperture at Keck is further divided into two
subapertures, with a baseline of 5 m, which are perpendicular to the
long baseline.  This results in a second, broader set of interference
fringes, perpendicular to the primary fringes, which are modulated to
take sky background measurements.  A detailed description of this
procedure can be found in \citet{koresko06}. 

Each individual observation of a star consists of about 10 minutes of
data collection, made up of several hundred null/peak (i.e.,
destructive/constructive) measurements. These measurements are
averaged to produce a final value for the null. Observations of
science objects are alternated with observations of calibrator stars.
Table \ref{tab-kinobs} lists the observations taken at KIN.  Further
details of the KIN design, observation procedures, and data reduction
can be found in \citet{colavita06}, \citet{koresko06}, 
\citet{colavita08} and \citet{colavita09}).

\begin{deluxetable}{ccccc}
\tablecaption{Summary of Observations at KIN \label{tab-kinobs}}
\tablewidth{0pt}
\tablehead{ \colhead{Target} & \colhead{Date} & \colhead{Start Time} &
  \colhead{\# Scans} & \colhead{Total duration} \\
\colhead{} & \colhead{} & \colhead{(UT)} & \colhead{} & \colhead {(s)}}
\startdata
$\beta$ Leo & 18 Feb 2008 & 10:27 & 3 & 1827\\
$\beta$ Leo & 16 Apr 2008 & 6:36 & 3 & 1827\\
{\it o} Leo & 16 Feb 2008 & 10:51 & 2 & 1351\\
{\it o} Leo & 17 Feb 2008 & 10:40 & 3 & 2157\\
{\it o} Leo & 14 Apr 2008 & 7:10 & 3 & 1867\\
\enddata
\end{deluxetable}

\subsection{Spitzer}

\begin{deluxetable*}{rllll}
\tablewidth{0pt}
\tablecaption{{\it Spitzer} Observations of $\beta$ Leo\label{spitzer_obs_tab}}
\tablehead{ 
\colhead{AOR Key} & \colhead{Date} & \colhead{Instrument} & \colhead{Module} &
\colhead{Integration} 
}
\startdata
3921152    & 2004 May 17  &IRAC Mapping & Ch 1-4  &   26.8 s $\times$ 5 dithers \\
18010112   & 2006 Dec 30  &IRAC Mapping & Ch 2,4  &   26.8 s $\times$ 5 dithers \\
4929793    & 2005 Jan 3   & IRS Staring & SL, LL  &   6 s $\times$ 1 cycle \\
14500608   & 2006 Jun 8   & MIPS Photometry & 24 \micron, 5 cluster pos. & 3 s $\times$ 4 cycles\\
14496768   & 2006 Jan 14  & MIPS Photometry & 24 \micron, 5 cluster pos. & 3 s $\times$ 4 cycles\\
9807616    & 2004 Jun 1   & MIPS Photometry & 70 \micron, default scale & 10 s $\times$ 3 cycles\\
4313856    & 2004 Jun 1   & MIPS Photometry & 70 \micron, fine scale & 3 s $\times$ 1 cycle\\
           &              &                 &  (9 cluster pos.)  &     \\ 
8935936    & 2004 Jun 1   & MIPS Photometry &160 \micron, 9 cluster pos. &  3 s
$\times$1 cycle \\
17325056   & 2006 Jun 12  & MIPS SED-mode   &  1\arcmin~chop  &  10s $\times$10 cycles \\
\enddata 
\end{deluxetable*}

$\beta$ Leo was observed using all three instruments on {\it
Spitzer~}: InfraRed Array Camera (IRAC, \citealt{fazio04}), InfraRed
Spectrograph (IRS, \citealt{houck04}) and Multiband Imaging Photometer
for {\it Spitzer} (MIPS, \citealt{rieke04}). Details about the
observations are listed in Table \ref{spitzer_obs_tab}. 
% A MIPS 160 \micron\ observation was also taken, but the integration was very
% shallow and no suitable Point Spread Function (PSF) observation was
% available for the 160 \micron\ filter leak subtraction. As a result,
% the 160 \micron\ data were not used. 
The observations at 24 \micron\
were done at two epochs (2006 Jan 13 and 2006 Jun 08) in standard
small-field photometry mode with four cycles with 3 s integrations at
5 sub-pixel-offset cluster positions, resulting in a total integration
of 902 s on source for each epoch. The 70 \micron\ observations were
done in both default- and fine-scale modes on 2004 May 31 with a total
on-source integration of 250 s for the default scale and 190 s for the
fine scale. The MIPS SED-mode observations were obtained on 2006 Jun
12 with 10 cycles of 10 s integrations and a 1\arcmin~chop distance
for background subtraction, resulting in a total of 600 s on
source. The 160 \micron\ observations were obtained with the original
photometry mode with an effective exposure of $\sim$150 s near the
source position. All of the MIPS data were processed using the Data Analysis
Tool \citep{gordon05} for basic reduction (e.g., dark subtraction,
flat fielding/illumination corrections), with additional processing to
minimize instrumental artifacts
\citep{engelbracht07,gordon07,lu08,stansberry07}. 
After correcting these artifacts
in individual exposures, the final mosaics were combined with pixels
half the size of the physical pixel scale for photometry
measurements. The calibration factors used to transfer the
instrumental units to the physical units (mJy) are adopted from
\citet{engelbracht07} for 24 \micron; \citet{gordon07} for 70 \micron;
\citet{lu08} for MIPS-SED mode data, and \citet{stansberry07} for 160
\micron.  

The IRS spectral data were processed starting with the Basic
Calibrated Data (BCD) products from the SSC IRS pipeline S18.7.  To
maximize the quality of the final spectrum, we adopted the extraction
methods developed by the Formation and Evolution of Protoplanetary
Systems (FEPS: \citealt{meyer06}) and Cores to Disks (C2D:
\citealt{evans03}) Spitzer legacy science teams, and based in part on
the SMART software package \citep{higdon04}.  Full details of the
process are presented in \citet{bouwman08} and \citet{swain08}.  Here
we briefly summarize the important elements of the extraction
technique.

We begin with the droopres intermediate BCD product. Background
subtraction is accomplished by subtracting associated pairs of the 2D
imaged spectra from the two nodded positions along the slit.  This
also eliminates stray light contamination and anomalous dark current
signatures. Bad pixels were replaced by interpolating the values in
the surrounding 8 pixels. A 6.0 pixel fixed-width aperture was used in
the extraction.  We optimized the position of the extraction aperture
for each order by fitting a sinc profile to the collapsed and
normalized source profile in the dispersion direction.  The {\it irsfringe}
package\footnote{http://ssc.spitzer.caltech.edu/dataanalysistools/tools/irsfringe/}
was used to remove low-level fringing for wavelengths $>$20
\micron.

Our custom extraction relies on relative spectral response functions
(RSRFs) derived independently using stars from the FEPS legacy program
free of any thermal excess emission \citep{bouwman08, carpenter08}.
This RSRF assumes that the object is perfectly centered in the slit,
but slight order mismatches are evident in the extracted spectra
suggesting that the object was not centered. Such an offset can induce
a wavelength dependent curvature in the extracted spectra because of
the fixed slit size and diffraction-limited PSF.  We employed an
algorithm developed by J. Bouwman \& F. Lahuis (described in
\citealt{swain08}) to correct for this offset.  This process
dramatically reduced the order offsets.  However, a small residual
offset remains between the SL1 and LL2 modules.

The final absolute flux density calibration is derived by processing
the SSC primary IRS calibrators with the exact same algorithms as used
for the FEPS stars \citep{carpenter08} and our program star.  The
uncertainty in the final absolute calibration is estimated to be
$\sim$10\%.

$\beta$ Leo was observed with IRAC at two epochs: 2004 May 7 (AOR Key
3921152 and 4306176); and 2006 Dec. 30 (AOR Key 18010112). In both the
3921152 and 18010112 datasets, the star was observed in full-frame
mode with 30 s frame times and a 5-position small-scale dither
pattern, for a total of 134 s integration time on-source in each
band. While the 3921152 dataset contains data in all IRAC bands,
images only at 4.5 and 8.0 $\mu$m were acquired in 18010112. In the
4306176 dataset, the star was observed in all bands in full frame mode
with a 12 s frame time, repeated twice and with no dithers. Due to the
much shorter exposure and the absence of dithering (limiting the
spatial sampling of the star on the IRAC arrays), we have discarded
the 4306176 dataset. For the other two datasets we first created
individual mosaics in each band starting from the BCD (pipeline
version S18.5.0), using our custom post-BCD software IRACproc
\citep{schuster06}, which is based on the SSC mosaiker MOPEX.

%{\bf IRAC reduction and flux estimation... (Massimo's description)
%{\it $\beta$~Leo is
%  saturated in the IRAC data, so the photometry was obtained by
%  fitting the PSF wing by Massimo Marengo. }}

\section{Results}
\label{results}

\subsection{BLINC}

\begin{figure}
\figurenum{1}
\label{fig-mmtnulls} 
\plotone{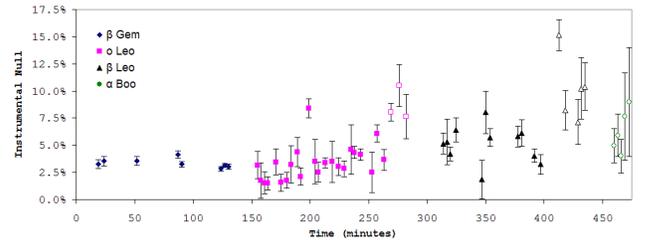} 
\caption{Instrument nulls of individual datasets for calibrator stars $\beta$ Gem (diamonds) and $\alpha$ Boo (circles) as well as for science stars {\it o} Leo (squares) and $\beta$ Leo (triangles) as a function of time. Hollow points indicate
  sets not used in our analyses due to weather. Error bars are
  determined by variations of the null within the set.} 
 
\end{figure} 

As an initial test for dataset quality, a 16$\times$20 pixel box centered
around the star (which corresponds to 0\farcs8$\times$1\arcsec~on the
sky) is used to calculate the BLINC instrument null. An elongated box is
used instead of a circular aperture because the aperture for BLINC
is elliptical, making the PSF of a star elliptical as well. The
16$\times$20 pixel box size ensures most (though not necessarily all)
of the flux from the system is measured. Figure \ref{fig-mmtnulls}
shows the average null for each set. Since each set contains either 10
or 40 frames, error bars are determined by the deviation of the nulls
in individual frames from the average null of the whole set. A filled
point is considered good data that is used for further analysis, while
a hollow data point is rejected due to weather. Weather-rejected data
occurred consistently for all stars when the telescope rotated into a
substantial wind. This wind resulted in problems maintaining the
telescope's AO system stability and, consequently, the null. 

For the good datasets, the calibration star $\beta$ Gem has only small variations both within a set
(as indicated by the small error bars) and from set to set. The fact that 
$\beta$ Gem has a non-zero null demonstrates the necessity of having 
calibrator stars: minor movements of the star due to vibrations in the 
system and sky turbulence result in imperfect suppression (chromatic effects
on the null are minimized by a ZnSe corrector in the non-shifted beam path). 
For the science stars, there is greater variation than is seen for $\beta$ 
Gem. This is primarily due to the
science stars being much dimmer in the N-band, making them more
sensitive to changes in background sky brightness. Using weighted
means to combine all the good measurements for each star, $\beta$ Gem
has an instrumental null of 3.19$\pm$0.09\%, {\it o} Leo has
3.33$\pm$0.28\% and $\beta$ Leo has 4.93$\pm$0.29\%. Using the $\beta$ Gem 
instrumental null as a baseline, we find source nulls of 0.14$\pm$0.30\% for 
{\it o} Leo and
1.74$\pm$0.30\% for $\beta$ Leo. This indicates that, within errors,
{\it o} Leo does not have resolved emission, while $\beta$ Leo has
resolved emission at 5$\sigma$ significance.

\begin{figure}
\figurenum{2}
\plotone{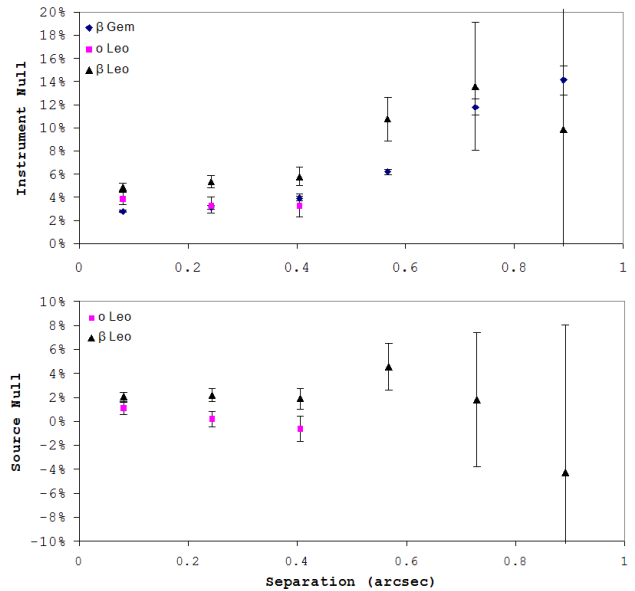}
\label{fig-annNulls}
\caption{Strip nulls as a function of separation from the center for
  $\beta$ Gem (diamonds), {\it o} Leo (squares) and $\beta$ Leo
  (triangles). (a) The instrument null. (b) The calibrator-subtracted 
  source null. Values and errors are determined from weighted
  means.}
\end{figure}

Information pertaining to the variation of the null with respect to
separation from the star can be gained by measuring the null within
strips of increasing vertical separation from the center of the PSF. Due to 
the vertically changing transmission pattern of BLINC (illustrated in Figure 
\ref{fig-TFs}a), such strips are more natural to use than circular annuli. 
Each dataset has its instrumental null calculated within 16$\times$3 pixel 
strips. For each separation, all good datasets for a star are
combined via weighted means, which also provides the errors. The
instrumental nulls for $\beta$ Gem, {\it o} Leo and $\beta$ Leo are
shown in Figure \ref{fig-annNulls}a, while Figure \ref{fig-annNulls}b
shows the source null for {\it o} Leo and $\beta$ Leo. The x-axis is
expressed as an angular separation from the center of the system,
since 1 pixel has a width of 0\farcs054.

In Figure \ref{fig-annNulls}a, both $\beta$ Gem and $\beta$ Leo show
an increase in the instrumental null beyond $\sim$0\farcs5. This is
likely caused by residuals from the adaptive optics systems, as well
as imperfect beam alignment (both of which are also evident in the
fact that the calibrator star has a non-zero null). Data for {\it o}
Leo only extend to 0\farcs4 since beyond this separation, slow
variations in sky background became comparable to the brightness of
{\it o} Leo itself when being imaged destructively. As a result, the
destructive signal became lost in the noise.

The source null of {\it o} Leo (Figure \ref{fig-annNulls}b) does not
show evidence (at a 3 $\sigma$ confidence or greater) of any resolved
emission within 0\farcs4. For $\beta$ Leo, however, an excess null is
present out to a separation of at least 0\farcs6 (beyond which the
error bars become too large to determine whether or not extended
emission is present). It must be cautioned that while Figure
\ref{fig-annNulls} provides evidence of extended emission, it cannot
be straightforwardly interpreted to be evidence for continuous emission
from the inner limit of BLINC out to at least 0\farcs6: the measured
excess is a result of the incoming excess multiplied by the
interferometer transmission function and convolved with the PSF of
the telescope aperture. Modeling is necessary to interpret the
figure, and is presented in section 4.3.2. 

\subsection{KIN}

Table \ref{tab-kinnull} shows the calibrated nulls derived from the
KIN observations.  The nulls in the table are an average of the
observations taken on each night, and are analogous to the
"source nulls" derived for the BLINC data, with both having been calibrated
using a calibrator star. The quoted errors are taken
to be the larger of the 'external' error for that date, or the
average of the 'formal' errors for all the scans on that date.  The
'formal' error is defined to be the scatter in the null within a given
scan, while the 'external' error is calculated as the standard
deviation among all scans on a given night, divided by the square
root of the number of scans (additional information on these errors
can be found in \citet{colavita09}).

\begin{deluxetable}{ccccc}
\tablecaption{Calibrated Nulls at KIN \label{tab-kinnull}}
\tablewidth{0pt}
\tablehead{
\colhead{Target} &
\colhead{Date} &
\colhead{Start Time (UT)} &
\colhead{Null (\%)} &
\colhead{Err. (\%)}
}
\startdata
$\beta$ Leo & 18 Feb 2008 & 10:27 & 0.9 & 0.7\\
$\beta$ Leo & 16 Apr 2008 & 6:36 & -0.1 & 0.3\\
{\it o} Leo & 16 Feb 2008 & 10:51 & -0.4 & 1.0\\
{\it o} Leo & 17 Feb 2008 & 10:40 & 3.6 & 0.3\\
{\it o} Leo & 14 Apr 2008 & 7:10 & 1.7 & 1.0\\
\enddata
\end{deluxetable}

\subsubsection{$\beta$ Leo}

We find that there is no significant nulled flux around $\beta$ Leo,
in contrast to the MMT results.  Implications of this result are
discussed in Section \ref{interferometric_model_bLeo}.  The quality
of data taken in the 2008 February run was low compared to the data taken
in the later run, resulting in larger errors.  Due to the relatively
low data quality, the 2008 February data were reduced 'manually' (i.e.,
taking into account the relatively low percentage of data passing
quality gates as well as some non-standard observing procedures),
rather than using an automated pipeline.  However, it should be noted
that the result of both sets of reductions (manual and pipeline) agree
within the errors.  Furthermore, both the February and April data are
consistent and suggest no resolved circumstellar emission within the
KIN aperture. 

\subsubsection{{\it o} Leo}

For {\it o} Leo, two of three sets of observations show evidence for
resolved flux.  The first set shows a negative null, which is
non-physical, though the result is consistent with zero resolved
emission within the errors.  The last set of data, taken on 14 April,
shows a non-zero null; however, large errors result in this detection
being only at a level of 1.8 $\sigma$, which we do not
consider to be significant.  The observations taken on 2008 February
17 yield a detection of
resolved emission that can be considered 
robust. This is in contrast to the MMT results which show no resolved
emission. However, interpretation of the KIN resolved emission is more
complicated as {\it o} Leo is a double-lined spectroscopic binary \citep{H01}. 
Detailed modeling (Appendix B) has determined that the resolved flux 
is consistent with emission from the stellar pair, which has
a separation of about 4.5 mas \citep{H01}.  
Furthermore, the behavior of the null over time is consistent with the
orbital solution calculated by \citet{H01}. If binarity is indeed the
source of the excess seen by KIN, this explains why it was not
observed by BLINC as well: the separation between the stars is much too 
small to be detected by BLINC. Additional information on {\it o} Leo can 
be found in the appendix. 

\subsection{{\it Spitzer} Results} 

The photometry at 24 \micron\ was obtained using aperture photometry
on each of the five cluster data sets for both epochs. Two aperture
settings were used (small: a radius of 6\farcs23 with sky annulus
between 19\farcs92 and 29\farcs88 and an aperture correction of 1.699;
large: a radius of 14\farcs94 with sky annulus between 29\farcs88 and
42\farcs33 and an aperture correction of 1.142). Such
aperture-corrected MIPS photometry generally agrees to within
$\sim$1\% for a clean, non-saturated point source. The large aperture
gives a total integrated flux density of 1623$\pm$33 mJy (2\% error
assumed), which is $\sim$2.5\% higher than the small one (This
behavior is similar to that of the resolved disk of $\gamma$~Oph at 24
\micron\ \citep{S08}). This result suggests $\beta$~Leo is slightly
more extended than a true point source at 24 \micron. Therefore, we
adopt the large aperture result as the final photometry in the 24
\micron\ band.

However, the FWHM measurements\footnote{Based on a 2D Gaussian fitting
function on a field of 103\farcs4.} of the source at 70 \micron\ are
consistent with it being a point source as compared to calibration
stars that have a similar brightness at this wavelength. The 70
\micron\ photometry was extracted using PSF fitting, giving a total
flux density of 711$\pm$50 mJy (7\% error assumed) in the 70 \micron\
band. The MIPS SED-mode data were extracted using an aperture of 5
native pixels ($\sim$50\arcsec) in the spatial direction. The final
MIPS SED-mode spectrum was further smoothed to match the resolution at
the long-wavelength portion of the spectrum (R=15).

The 160 \micron\ observation needs additional attention to eliminate
the filter leak contamination. The expected stellar photospheric
brightness at 160 \micron\ is $\sim$26 mJy, resulting in a leak strength
of $\sim$390 mJy ($\sim$15 times) that is brighter than the expected
disk brightness assuming its emission is blackbody-like. Visual
inspection of the 160 \micron\ data confirms that the $\beta$ Leo
observation was mostly dominated by the ghost image (off
position). We used the 160 \micron\ observation (AOR Key 15572992 from
PID 52) of Achernar (HD10144, B3Ve) for the leak subtraction as both
160 \micron\ observations were obtained in the same way (dithered cluster
position). A constant background value of 7.5 MJy~sr$^{-1}$ for
Achernar and of 10 MJy~sr$^{-1}$ for beta Leo was taken out in each of
the mosaics first. We then scaled the Achernar data by a factor of
0.34 for subtraction, and a faint source appeared in the expected
position of $\beta$ Leo after the subtraction. Because the background
(most due to instrument artifacts) is not very uniform, the source is
only weakly detected. The pixel-to-pixel variation of the observation
suggests a point source 1-$\sigma$ sensitivity of 30 mJy in this observation.  
We placed a 16\arcsec~radius aperture at the position of the source to 
estimate the source brightness. After proper aperture correction
\citep{stansberry07}, a flux density of 90 mJy is estimated.  
Achernar is a nearby (42.75 pc; \citealt{vanleeuwen07}) fast rotating
Be star with a close (0\farcs15) companion \citep{kervella07}. 
\citet{kervella08} estimated the companion has spectral type of
A1V-A3V based on the near-infrared colors and contributes 3.3\% of
the combined photospheric emission. The color-corrected {\it IRAS}
flux densities are consistent with the expected levels of the
photosphere represented by the Kurucz model of $T_{\ast}\sim$15,000 K,
$log g$=3.5, and solar metallicity \citep{kervella09}, scaled to match
the distance of the star and stellar radius of 8.5 $R_{\sun}$. Using
the same photospheric model, the expected 160 \micron\ value for
Achernar is 68 mJy. Part of Achernar's photosphere is
subtracted off from the $\beta$ Leo data for the leak
subtraction. After correcting this, the faint source at the expected
$\beta$ Leo position has a flux density of 114 mJy. However, this
value is a lower limit due to the uncertainty in scaling the leak
subtraction. Varying the scaling by $\pm$10\% results in a 30\% change in 
the photometry. Because of the difficulty of leak subtraction and low
signal-to-noise 
of the data, we do not consider the source to be detected at 160
\micron. The source brightness is likely within 114--217
mJy. Futhermore, {\it Herschel} has measured $\beta$~Leo with PACS at
100 and 160 \micron\ with flux densities of 500$\pm$50 mJy and
230$\pm$46 mJy, respectively \citep{matthews10}. We then adopted the
{\it Herschel} values to constrain the amount of cold dust in the
$\beta$ Leo system in later analysis.

\begin{figure}
\figurenum{3} 
\label{sed} 
\plotone{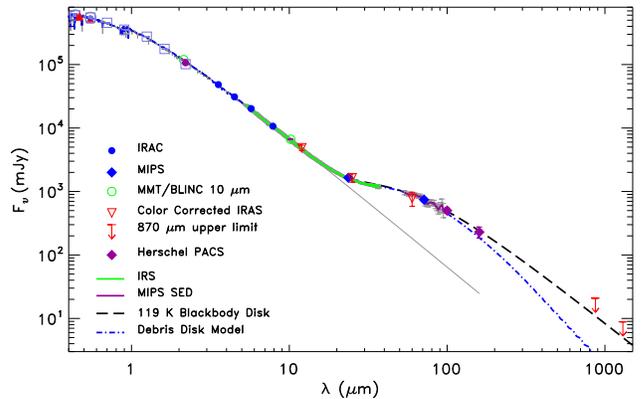}
\caption{The SED of the $\beta$~Leo system. Most of the symbols and lines
  used are shown on the plot except for the optical ground-based
  photometry (open squares from Johnson system, filled circles from
  Hipparcos Tycho system, filled triangle from Str\"omgen system). The
  870 \micron\ upper limit (3$\sigma$) is from \citet{holmes03}.} 
\end{figure}

Since $\beta$ Leo is saturated in the IRAC observations at all bands,
we extracted the photometry based on PSF fitting. We have derived its
Vega magnitudes by fitting the unsaturated wings and diffraction
spikes with a PSF derived from the observations of bright stars (Vega,
Sirius, $\epsilon$ Eridani, Fomalhaut, and $\epsilon$ Indi). The
construction of this PSF and the details of the adopted PSF-fitting
technique are described in \citet{marengo09}. The PSF files are
available at the SSC web
site\footnote{http://ssc.spitzer.caltech.edu/irac/psf.html}.  We
estimated the magnitude uncertainty by bracketing the best fit with
over- and under-subtracted fits in which the PSF-subtraction residuals
were comparable to the background and sampling noise. The IRAC
photometry for $\beta$ Leo based on the epoch 1 data is magnitudes
(Vega) of 1.905 $\pm$ 0.011, 1.909 $\pm$ 0.011, 1.875 $\pm$ 0.011, and
1.927 $\pm$ 0.011 for the IRAC [3.6], [4.5], [5.6]. and [8.0] bands,
respectively. The data quality in the second epoch is lower than in
the first one, so we used it only as a rough confirmation of these
results.  We converted the PSF-fit magnitudes into fluxes by adopting
the zero point (Vega) fluxes listed in the IRAC Data Handbook version
3.0 (2006).  The resulting photometry and associated errors are listed
in Table 5.  The spatial extension of the $\beta$ Leo disk could not
be constrained with the IRAC images because of the PSF core
saturation. The measured photometry and spectra along with previous
measurements 
from the literature are shown in the spectral energy distribution
(SED) in Figure \ref{sed}. 

\begin{deluxetable*}{crrrrrrll}
\tablewidth{0pt} 
\tablecaption{Broadband Phometry\tablenotemark{a} of $\beta$ Leo\label{tab_phot}}
\tablecolumns{7}
\tablehead{ \colhead{$\lambda_c$ [\micron]} & \colhead{Star $F_{\nu}$
    [Jy]} & \colhead{Total $F_{\nu}$ [Jy]}  & \colhead{Error $F_{\nu}$
    [Jy]} & \colhead{$\chi$\tablenotemark{b}}&
  \colhead{Fraction [\%]} & \colhead{Origin}
}
\startdata
   $[3.6]$  &  47.967  &  48.435  &   1.085  &     0.4  &     1.0  &      IRAC  \\ 
   $[4.5]$  &  31.083  &  30.971  &   0.693  &    -0.2  &    -0.4  &      IRAC  \\ 
   $[5.8]$  &  19.731  &  20.450  &   0.458  &     1.6  &     3.6  &      IRAC  \\ 
   $[8.0]$  &  10.547  &  10.759  &   0.244  &     0.9  &     2.0  &      IRAC  \\ 
  23.675  &   1.189  &   1.647  &   0.033  &    13.9  &    38.5  &      MIPS  \\ 
  25.000  &   1.066  &   1.647  &   0.247  &     2.4  &    54.6  &  IRAS  \\ 
  71.419  &   0.128  &   0.743  &   0.052  &    11.8  &   481.6  &      MIPS  \\ 
\enddata
\tablenotetext{a}{Photometry longward of 10 \micron\ is color-corrected.} 
\tablenotetext{b}{Significance of the excess defined as
  $\chi= (F_{Total}-F_{Star})/F_{Error}$.}
\label{photometric_observations}
\end{deluxetable*}

% Table \ref{photometric_observations}
% summarizes the photometric observations of $\beta$ Leo.

\section{Analyses}
\label{analyses}

In this section, we use the measurements described above to
characterize the infrared excess emission of $\beta$ Leo. We first
establish the properties of the star, which we use to extrapolate the
photospheric output from the near-infrared. This process lets us
subtract the photospheric contribution to the mid-infrared
measurements to reveal the excess due to circumstellar dust.  We then
consider the upper limit on the extended emission from the KIN.
Finally, we simulate the BLINC data to determine the constraints it
places on the disk structure at 10 $\mu$m.

\subsection{Stellar Photospheric Emission}
\label{stellar_photosphere} 

To determine the stellar photospheric emission, we fit all available
optical to near-infrared photometry (Johnson $UBV$, Str\"omgen $uvby$
photometry, Hipparcos Tycho $BV$ photometry, 2MASS $JHK_s$ photometry)
with the synthetic Kurucz model \citep{castelli03} based on a $\chi^2$
goodness of fit test. Because of the star's proximity, the 2MASS
photometry is saturated and unusable. In addition, $\beta$~Leo has
been found to have a hot $K$-band (2.14 \micron) excess of
2.17$\pm$1.4 Jy \citep{akeson09} ($\sim$2.5\% above the photosphere).
To overcome these difficulties, we tried to include other ground-based
near-infrared photometry. 
% From the UKIRT standard listing, $\beta$ Leo is at $J$=2.02 mag, $H$=1.99 mag, and
% $K$=1.98 mag with errors $\leqq$ 0.01 mag.  
$\beta$ Leo is a primary bright standard in the
UKIRT/Mauna Kea System, with m$_K$ = 1.98, $H-K$ = 0.01, and $J-K$
=0.04\footnote{available at the URL
  http://www.jach.hawaii.edu/UKIRT/ astronomy/calib/phot\_cal/bright\_stds.html}.   
% We transferred additional
% Johnson $K$-band photometry to the 2MASS system, and also took out the
% contribution of the hot excess seen at 2.14 \micron\ . Combining all
% the available data 
Combining all the available data and transforming to the 2MASS system 
resulted in a $K_s$ of 1.93 mag for the photospheric fitting.

A value of $T_{eff}$=8500 K with a total stellar luminosity of 13.44
$L_{\sun}$ integrated over the Kurucz model spectrum at the given
distance gives the best $\chi^2$ value. This suggests that the stellar
radius is 1.7 $R_{\sun}$ using the Stefan-Boltzmann equation. In
comparison, \citet{akeson09} estimate the stellar radius to be
1.54$\pm$0.021 $R_{\sun}$ based on the long-baseline interferometric
observation from CHARA array. Their radius was derived including a fit
with an incoherent flux (hot excess), while the early VLTI interferometric
observation by \citet{difolco04} derived a much larger radius
(1.728$\pm$0.037 $R_{\sun}$) without including the incoherent
flux. Additionally, \citet{decin03} use various methods to determine a
stellar radius in the range of 1.68 -- 1.82 $R_{\sun}$ for
$\beta$~Leo, consistent with a 
spectral type of A3V. A stellar radius of 1.5 $R_{\sun}$ is the
nominal value for a F0V star with an expected effective temperature of
7300 K, making it too cool to fit the observed photometry for
$\beta$~Leo. We do not attempt to solve the mystery of stellar radius
for $\beta$ Leo. However, we use 1.7 $R_{\sun}$ as the stellar radius
in the thermal models for computing dust temperatures in order to be
consistent with the stellar radiation field. (The difference in
stellar radius produces a $\sim$22\% difference in input stellar
luminosity). Using a stellar mass of 2.0 $M_{\sun}$ and the derived
stellar luminosity, the blowout size ($a_{bl}$ in radius) is $\sim$3
\micron\ for astronomical silicate grains \citep{laor93}.

Based on the best-fit Kurucz model, we then estimate the stellar
photospheric flux densities at the wavelengths or bands of
interest. For MIPS observations, the expected photospheric flux
densities are computed based on the monochromatic wavelength of each
band, while the expected photospheric fluxes are integrated over the
bandpasses for the IRAC observations. The additional 2\% errors of the
photosphere determination (mostly limited by the K-band photometry
accuracy) have been added to the IRAC
measurements. The broadband phometry obtained by this work and their
corresponding photospheric flux densities are listed in Table
\ref{tab_phot}. 
% along with the excess
% levels are listed in Table \ref{tab_phot}.

\subsection{Infrared Excess around $\beta$ Leo}
\label{sec-Spitzermod}

\subsubsection{Near Infrared}

% $\beta$ Leo is a primary bright standard in the
% UKIRT/Mauna Kea System, with m$_K$ = 1.98, $H-K$ = 0.01, and $J-K$
% =0.04\footnote{available at the URL
%   http://www.jach.hawaii.edu/UKIRT/astronomy/calib/phot\_cal/bright\_stds.html}. 

The colors of beta Leo in the UKIRT/Mauna Kea System are $H-K$ = 0.01,
and $J-K$ = 0.04 with estimated errors of 1\%. 
They agree nearly perfectly with 
expectations for the spectral type of the star, A3V
\citep[e.g.,][]{tokunaga00}.  The IRAC measurements at [3.6]
and [4.5] are also in agreement with those for an A3V star, to within
their errors of $\sim$ 1\%. Thus, there is no evidence for an
excess at the 2.5\% level as suggested by the results of \citet{akeson09}, 
unless the excess has near infrared colors identical
to those for $\beta$ Leo itself. Nevertheless, these photometry
results cannot rule out any hot 2 \micron\ excess at the 1\%
level. New observations obtained with CHARA/FLUOR suggest that the
K-band excess around $\beta$ Leo could be smaller than what was reported in 
\citet{akeson09} (O. Absil priv. comm.).

\subsubsection{{\it Spitzer} Mid- and Far-Infrared Photometry}

The expected photospheric contribution in the MIPS 24 \micron\ band is
1190 mJy, indicating that the flux seen at 24 \micron\ is dominated by
the star and that the excess emission is 433 mJy (before color
correction). However, the flux seen in the 70 \micron\ band is
dominated by the excess emission (583 mJy before color correction)
compared to the stellar photosphere. This suggests that the dust
emitting at MIPS wavelengths has a color temperature of $\sim$120
K. Therefore, color correction factors of 1.055 and 1.054 are applied
to the excess emission at 24 and 70 \micron, respectively. The
infrared excess flux densities are then 457 and 615 mJy in the 24 and
70 \micron\ bands with assumed calibration errors of 2\% and 7\%,
respectively.

\begin{figure}
\figurenum{4} 
\label{mips24im} 
\plotone{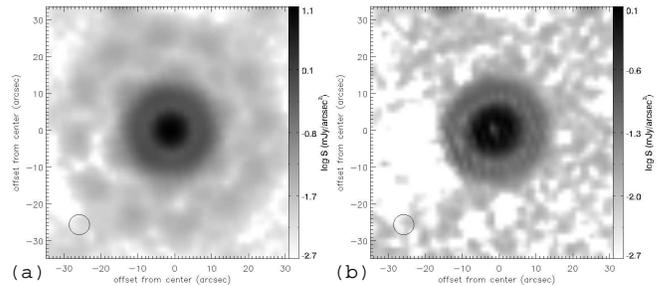}
\caption{MIPS 24 \micron\ images of the $\beta$ Leo disk with N up and E
toward the left. The nominal beam size (FWHM) is indicated as black
circles on the images. The surface brightness scale is to the right of
each image. (a) 24 \micron\ image after photospheric subtraction. (b)
24 \micron\ image after over-subtraction of the model stellar
photosphere. }
\end{figure}

\begin{figure}  
\figurenum{5}
\label{radprof24}
\plotone{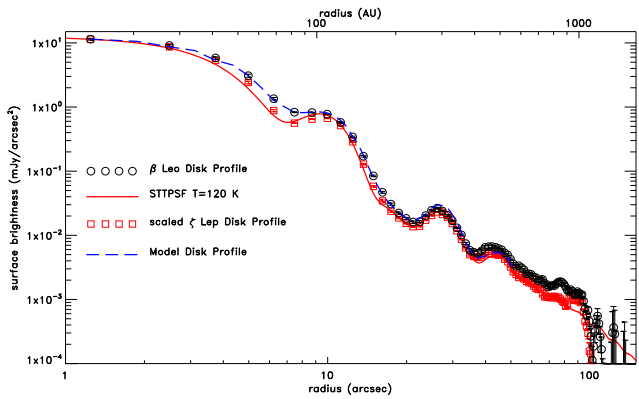}
\caption{The azimuthally averaged radial profile of the $\beta$~Leo
  disk at 24 \micron\ shown in open circles compared to the model
  profile (blue long-dashed line). For comparison, the profiles made
  from point sources (STinyTim PSF and unresolved $\zeta$~Lep disk) are
  shown as the red solid line and open squares, respectively.} 
\end{figure}

At 24 \micron, the stellar photosphere was subtracted from the data by
scaling an observed point spread function (PSF) to the expected
photospheric flux density of $\beta$~Leo and to 1.25 times this
value. This over-subtraction technique has been applied to other 24
\micron\ resolved disks like $\gamma$~Oph \citep{S08} and Fomalhaut
\citep{S04} to reveal excess emission lying close to the star. The
results are shown in Figure \ref{mips24im}.  The FWHM\footnote{Based
on a 2D Gaussian fitting function on a field of 26\farcs1.} of the 24
\micron\ image is 5\farcs77$\times$5\farcs75 before the photospheric
subtraction, but is 6\farcs88$\times$6\farcs61 after the subtraction,
which is considerably larger than a typical red PSF (for example, the
$\zeta$~Lep disk has a FWHM of 5\farcs61$\times$5\farcs55,
\citealt{S08}). The resolved core emission at 24 \micron\ is best
shown in the average radial surface brightness profile (Figure
\ref{radprof24}).  It is evident that the first dark Airy ring
(between radii of 5\farcs5--7\farcs5) in the $\beta$~Leo profile is
filled in compared to a true point source, and the profile matches
well with that of a point source at larger radii.  The position angle
of the disk (after photospheric subtraction) is 118\arcdeg~with an
error of 3\arcdeg~estimated from two epochs of data. The ratio (1.041)
between the semi-major and semi-minor radii in the FWHM suggests the
disk is viewed at 20\arcdeg~$\pm$10\arcdeg~, close to face-on (a ratio 
of 1.011 is expected for a
point source while a ratio of $\sim$1.115 is derived from the
$\gamma$~Oph disk with inclination of $\sim$50\arcdeg). The low
inclination angle of the disk is consistent with the star being
inclined at 21.5\arcdeg~from pole-on \citep{akeson09}.

The excess flux densities at 24 and 70 \micron\ suggest a color
temperature of 119 K. A blackbody emission of 119 K represents the
excess emission (from the spectral shape of both the IRS and MIPS-SED data)
reasonably well, although the emission at 15--20 \micron\ is a
bit low (see Fig.~\ref{irs_excess}). For blackbody-like grains, a dust
temperature of 
$\sim$119 K corresponds to a radial distance of $\sim$20 AU in the
stellar radiation field of $\beta$~Leo, consistent with the disk being
marginally resolved at 24 \micron\ but not at 70 \micron. The total
dust luminosity is 1.2$\times$10$^{30}$ ergs s$^{-1}$ based on the 119
K blackbody radiation at the given distance of 11 pc, suggesting 
a dust fractional luminosity ($f_d$) of 2.4$\times$10$^{-5}$. 

\subsubsection{IRS Spectra}

\begin{figure}
\figurenum{6}
\label{irs_excess} 
\plotone{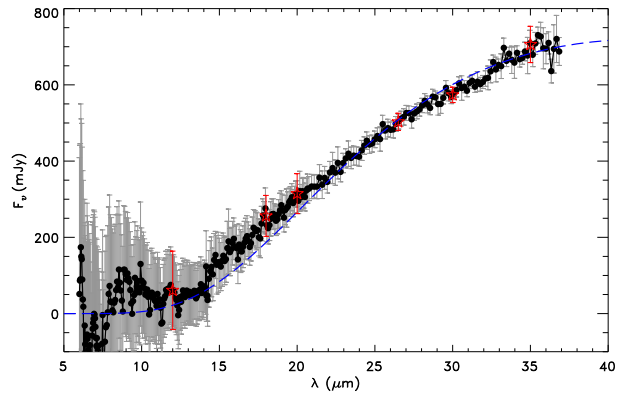}
\caption{Spitzer IRS spectrum of the excess emission around $\beta$
Leo (after stellar photospheric subtraction). The error bars include
additional 2\% errors from the photosphere determination. For
comparison, the long-dashed line is the blackbody emission of 119 K
normalized to the excesses at 24 and 70 \micron. The
star-symbols show the 6 fiducial points chosen in our model fitting
(see Sec \ref{model_main_belt} for details).}
\end{figure}

As described in Section 2.3, the flux offset between the IRS SL and LL
modules due to pointing was corrected using the automatic IRS
pipeline. A small residual offset is still evident when comparing the
slope of the observed SL spectrum with the slope of the expected
photosphere from the Kurucz model. We, therefore, scaled down the
extracted SL spectrum by 2.7\% so that the join points ($\sim$14.3
\micron) between SL and LL modules are smooth. The excess IRS spectrum
after photospheric subtraction is shown in Figure \ref{irs_excess}. At
12 \micron, the excess is $\sim$1.3\% above the stellar photosphere,
but $>$10\% for wavelengths longward of 18 \micron.

\subsection{Interferometric Analysis for $\beta$ Leo} 
\label{interferometric_model_bLeo}

Using the interferometric observations of {\it o} Leo, we were able to
put constraints on any dust that may be present in the {\it o} Leo
system and confirm the calibration that was used in the analysis of the 
$\beta$ Leo system. Details on the interferometric analysis of
{\it o} Leo are presented in Appendix B.

\subsubsection{KIN}

The KIN observations of $\beta$ Leo suggest that there is no
significant resolved emission at 10 \micron\ in the KIN beam. 
Limited by its small field of view, our KIN observations are unable to detect extended structures beyond a radius of $\sim$ 3 AU at the distance of $\beta$ Leo (11 pc). Moreover, due to its complex transmission function, KIN observations are most sensitive to extended structures with a radius of less than $\sim$1 AU. From the
errors during the April 2008 run, we adopt a 3 $\sigma$ upper limit of
0.9\% for the 10 $\mu$m resolved emission, relative to the stellar
photosphere. However, due to the complex transmission function of KIN 
(illustrated in Figure \ref{fig-TFs}b in the appendix), this limit cannot 
be directly converted into a flux. Instead, we use the Visibility Modeling Tool
(VMT) provided by the NASA Exoplanet Science Institute\footnote{available at 
the URL http://nexsciweb.ipac.caltech.edu/
vmt/vmtWeb/} to test different
excess configurations and 
determine how their simulated excesses compare to our limit. 

One possible component to any excess detected by KIN is the partially resolved photosphere of 
$\beta$ Leo itself. If we take the stellar radius to be 1.7 $R_{\sun}$ and 
assume the star to be uniformly bright, this excess only
amounts to 0.2\%. While this is well below our detection threshold, it is 
a constant source of excess that needs to be considered when evaluating models
that exceed our 0.9\% upper limit. 	

KIN is most sensitive to emission between $\sim$ 0.05 AU and 1 AU. A disk extending from 0 to 0.05 AU
would produce a null above the 0.9\% limit only if it were brighter than
645 mJy, whereas a ring between 0.05 and 0.1 AU would be detected
if its flux is above 145 mJy at 10 $\mu$m. 
A uniform disk model extending from 0 to 1 AU will produce an excess
greater than 0.9\% for 10 \micron\ fluxes above 110 mJy. However,
because the sensitivity drops rapidly at AU scales, a similar  
model extending from 0 to 2 AU will only surpass our 3 $\sigma$ limit if it is 
at least 240 mJy in brightness. For a 1 to 2 AU ring, the threshold becomes 370 
mJy.

% One component to any excess is the partially resolved photosphere of 
% $\beta$ Leo itself. If we take the stellar radius to be 1.7 $R_{\sun}$ and 
% assume the star to be uniformly bright, this excess only
% amounts to 0.2\%. While this is well below our detection threshold, it is 
% a constant source of excess that needs to be considered when evaluating models
% that exceed our 0.9\% upper limit. At small separations, KIN is best able to 
% detect structures extended beyond 0.05 AU: a ring between 0.05 and 0.1 AU would
% produce an null above the 0.9\% limit if it were brighter than 145 Jy, while 
% a disk extending from 0 to 0.05 AU would only be detected if it were above
% 645 Jy. 

% At AU scales, the VMT predicts that a uniform disk model extending from 0 to 1 
% AU will produce an excess greater than 0.9\% for 10 \micron\ fluxes above 110 
% mJy. However, because the sensitivity drops rapidly at AU scales, a similar 
% model extending from 0 to 2 AU will only surpass our 3 $\sigma$ limit if it is 
% at least 240 mJy in brightness. For a 1 to 2 AU ring, the threshold becomes 370 
% mJy. Thus, KIN can put the tightest limits on emission within $\sim$1 AU of 
% $\beta$ Leo.

The hot excess of $\beta$ Leo detected by \citet{akeson09} at 2 $\mu$m
is taken to be evidence for a hot inner ring. While there are several
models presented in \citet{akeson09} that can fit the hot excess, they
focus on a ring model extending from 0.13 to 0.43 AU. Such a ring
would be above KIN's 3 $\sigma$ limit if it has a 10 \micron\ excess
greater than 96 mJy. Based on their 2 \micron\ excess of 2.7 $\pm$ 1.4
Jy, a normal Rayleigh-Jeans falloff from this point would yield 108 mJy,
which is somewhat above our 3 $\sigma$ limit. Thus, the KIN null
detection is not consistent with the \citet{akeson09} result. There
are several possibilities that could explain the discrepancy. It is
possible that the excess is more compact than the suggested
model. However, to avoid being detected in KIN, the excess would have
to be closer to the star than 0.1 AU (if it is located within a narrow
ring) or 0.15 AU (if it is part of a continuous disk extending into
the star). In either case, much of the disk would be located within
the sublimation radius for amorphous grains of 0.12 AU
\citep{akeson09}, making the possibility of such a compact disk
unlikely. More likely, the disagreement between KIN and
\citet{akeson09} suggests that the source of this hot emission is
variable, has a
color bluer than a normal star (Rayleigh-Jeans), or the emission level
is smaller than the reported value. 

\subsubsection{BLINC} 
\label{Sec-BLINCmodel}

To compare physical models of the $\beta$ Leo disk to the BLINC data
(shown in Figure \ref{fig-annNulls}), a program was created to take an
input disk geometry, interfere it, convolve it with the aperture PSF
and pass it through the same data reduction pipeline used on real
images. Because of the limited datapoints, these calculations assumed
an optically and geometrically thin disk composed of blackbody
grains. The dust in the disk was given an optical depth of the form
\begin{equation} \tau (r) = Cr^{p}, \end{equation} \noindent where $C$
is a brightness scaling constant and $p$ is the power law
index. Assuming a sharp inner ($R_{in}$) and outer ($R_{out}$) cutoff
beyond which $\tau$ is zero, there were four variables: $R_{in}$,
$R_{out}$, $C$ and $p$.

\begin{figure} 
\figurenum{7}
\label{fig-BContMod}
\plotone{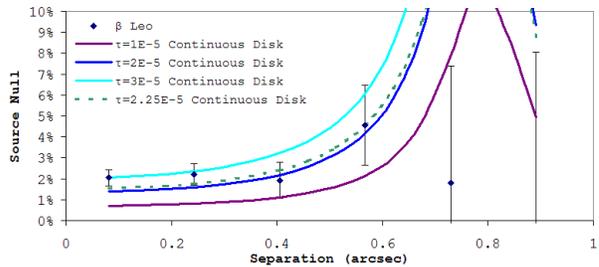}
\caption{Continuous disk models for varying brightness constants
plotted with $\beta$ Leo strip datapoints as a function of
separation from the center. The best fit continuous disk is indicated
with a dotted line. } 
\end{figure}

We first compare the possible disk geometries from the
  interferometry, using the integrated excess from the {\it Spitzer}
  observations as an additional constraint.
Figure \ref{fig-BContMod} shows a set of disk geometries compared to
the source null of $\beta$ Leo. In this and similar figures it should
be noted that the hump located near 0\farcs8 does not correspond to a
physical excess at that separation from the star; rather, it is a
feature relating to the PSF of the aperture. The disk geometries in
Figure \ref{fig-BContMod} use different brightness scaling constants,
$C$, and all assume a uniform ($p$=0) disk that extends from 0 AU to
10 AU. For a uniform disk, the values of $C$ are equivalent to the
optical depth of the disk. Disk geometries with $R_{out} > 10$ AU do
not differ significantly from these simulations, since the BLINC data
are not sensitive outside this radius. These models are thus good
approximations of continuous disks that extend into the outer regions
imaged by {\it Spitzer}. The best fit for such a 'continuous disk'
simulation is given as a dotted line in Figure \ref{fig-BContMod}, and
corresponds to an optical depth of 2.25\sci{-5}. This geometry has a
reduced $\chi$$^{2}$ of 4.2, making it a relatively poor fit. More
importantly, such a disk is required to have a flux density of 0.515
Jy at 10.5 \micron\ (or $\sim$9\% excess emission above the stellar
photosphere). This is inconsistent with the SED-measured excess shown
in Figure \ref{sed} and the IRS excess spectrum in Figure
\ref{irs_excess}. We therefore rule out a continuous disk extending
from 0 AU to 10 AU or beyond as being the source of the excesses
measured by BLINC.

\begin{figure} 
\figurenum{8}
\label{fig-BBestMod}
\plotone{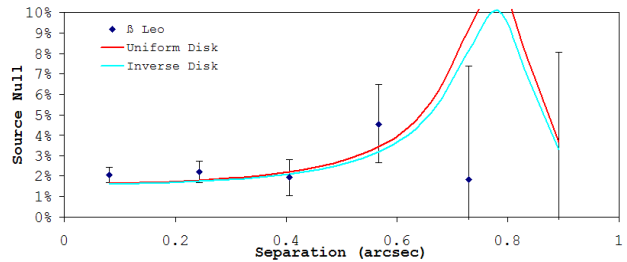}
\caption{Best fit models for a uniform disk and an inverse power law
disk, given a maximum flux of 0.3 Jy at 10.5 \micron. Also plotted 
are $\beta$ Leo datapoints, as a function of separation
from the center.}
\end{figure}

Better fits can be achieved by varying the
inner and outer radii. The best possible fits for both a uniform and
an inverse ($p=-1$) disk are with a ring extending from 1 to 2 AU. In both
models there is a reduced $\chi$$^{2}$ of around 0.7. However, these
models require the disk to be even brighter at 10.5 \micron\ than the
best fit continuous disk models, and thus are also inconsistent with
our other observations. In order to bring the models more in line with
the IRS data, we consider only those that are at most 0.30 Jy in
brightness ($\sim$5\% of the photosphere). Doing so, we find the best 
fit models to have reduced $\chi$$^{2}$ of around 2. The models are shown 
in Figure \ref{fig-BBestMod}. One model assumes a uniform disk and the other 
assumes an inverse disk. Both of these model disks have an inner radius of 2 
AU and a width of less than 1 AU (although the best fit uniform disk is 
slightly more extended than the best fit inverse disk). Both disks have a 
flux of 0.30 Jy at 10.5 \micron\, with the uniform disk having a slightly
better reduced $\chi$$^{2}$ ($\chi$$^{2}$ = 1.9) than the inverse disk
($\chi$$^{2}$ = 2.0). The uniform disk has an optical depth of
1.04\sci{-4} while the inverse disk has an optical depth of
1.50\sci{-4} at $R_{in}$.

Overall, both uniform and inverse models are similar with regards to
their quality of fit. The similarity of these two cases shows that the
best fit is insensitive to the power law index when considering thin
rings. 
%and concluding which disk (uniform or inverse) is the better fit
%cannot be done with this data. 
Also, as noted above, better fits can
be found if we allow for higher fluxes; however, such brighter disks
would be inconsistent with other data.

While the above models are the best fits to the BLINC data, $R_{in}$
and $R_{out}$ have some flexibility: $R_{out}$ can be extended to 4 AU
before the reduced $\chi$$^{2}$ exceeds 3. Similarly, disks that have
brightness less than 0.30 Jy and reduced $\chi$$^{2}$ less than 3
exist for models with $R_{in}$ as small as 1 AU. Reducing $R_{in}$ to
less than this extends the disk to regions where BLINC strongly
suppresses emission and is thus insensitive, resulting in a rapidly
increasing disk brightness while minimally affecting the quality of
fit. However, KIN becomes increasingly sensitive to regions inside of 
$\sim$2 AU, which restricts the amount of excess
that can be there and still result in a null detection by KIN. Disk
widths larger than 1.5 AU result in poor fits ($\chi$$^{2}$ 
greater than 3). However, disks narrower than 1 AU do not strongly 
degrade the fit; at widths less than $\sim$0.5 AU, the resolution of 
BLINC makes the models degenerate in brightness and width. 

With these possible geometries, we look in more detail at the flux
constraints from the interferometry. 
Based on the null analysis in Section 3.1, the excess was found to be
1.74$\pm$0.30\% over the photosphere. If we take the photosphere to be
6.3 Jy at 10.5 \micron, this excess would be 0.11$\pm$0.02
Jy. However, for a near face-on disk, the interference pattern will
suppress at least half of the light, and possibly more depending on
the location of the dust, which sets a lower limit of 0.22 $\pm$ 0.04
Jy on the dust emission in this ring. This is consistent with our
uniform and inverse model simulations that find that any disk with a flux fainter than 0.2 Jy would result in poor fits. Taking into
consideration the IRAC 8 $\mu$m and IRS data, the maximum brightness
for this BLINC component is 0.3 Jy.

We have used the VMT to model what level of signal, if any, KIN would
see from this belt. We find that a 0.3 Jy belt extending from 2 to 3
AU creates a null in KIN of -0.2\%. This negative null is likely due
to the complex transmission pattern of KIN (see Figure \ref{fig-TFs}b
in the appendix) and the fact that the belt would lie in the sidelobes
of this pattern. Combined with the partially resolved photosphere of
$\beta$ Leo, the net signal from the system is 0\%. For a belt that is
less than 0.3 Jy in brightness, the combined null approaches 0.2\% (as
we would expect). For a belt that has an inner radius as small as 1 AU,
the predicted null would still not be large enough to exceed the 0.9\% limit
that has been established. So, for belts less than 0.3 Jy, the resulting 
nulls predicted by the VMT are fully consistent with our KIN data. In
summary, using the simple models outlined above we then conclude that
the excess component detected by BLINC is consistent with a narrow
ring of 2--3 AU with a brightness of 0.25$\pm$0.05 Jy at 10.5 $\mu$m.

\section{Physical Models of the $\beta$ Leo Debris Disk}
\label{physical_model}

We now build a physical model to probe the disk properties
(such as the extension and total dust mass) based on all the information derived
from interferometric and {\it Spitzer} excesses, with assumed grain
properties. We start with a simple disk with one component in order to minimize
the free parameters being fit to the global excess SED; we then add an 
additional component required by the spatial constraint provided by
the interferometric data. 

\subsection{Basic SED Model Description} 

We assume a simple, geometrically-thin (1-D) debris disk where the
central star is the only heating source in the system (optically
thin). The dust is distributed radially from the inner radius
($R_{in}$) to the outer radius ($R_{out}$), and governed by a power
law of radius $r$ for the surface number density with an
index $p$ ($\Sigma(r) \sim r^p$, $p$ = 0 is a disk with constant
surface density expected from a Poynting-Robertson drag dominated
disk; while $p$ = --1 is a disk expected from grains ejected out of
the system by radiation pressure). We further assume that the grains in
the disk have a uniform size distribution at all radii, following a
power-law with minimum cutoff radius of $a_{min}$,
maximum cutoff radius of $a_{max}$, and a power index of $q$ = --3.5
($n(a)\sim a^{-3.5}$), consistent with grains generated in theoretical
collisional equilibrium. No obvious dust mineralogical features are
seen in the IRS spectrum to favor a specific dust composition;
therefore, we adopt the grain properties (size-dependent absorption
$Q_{abs}$ and scattering $Q_{scat}$) of astronomical silicates
\citep{laor93} with an assumed grain density of 2.5 g~cm$^{-3}$.  We
then compute the dust temperature as a function of 
the grain size, the incident stellar radiation (best-fit Kurucz
model), and the distance $r$ based on balancing the energy between
absorption and emission by the dust (scattering is ignored in this
simple model). The final SED is then integrated over the grain size
and density distribution.

\subsection{Main Planetesimal Belt}
\label{model_main_belt}

\begin{figure} 
\figurenum{9}
\label{sed_excess}
\plotone{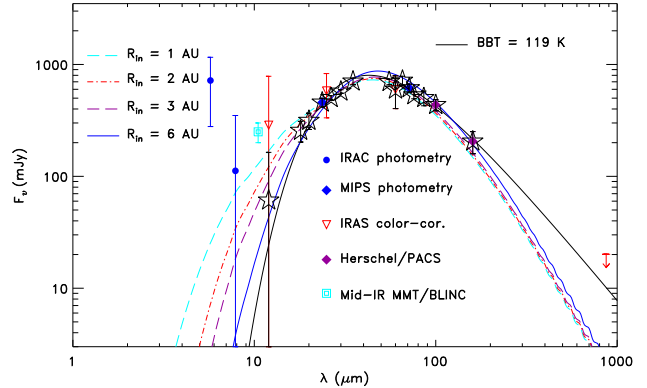}
\caption{The excess SED of the $\beta$~Leo disk. Most of the symbols and lines
used are defined on the plot, except for the open-star symbols which
are the 15 fiducial points used in the $\chi^2$ computation. }
\end{figure}

Figure \ref{sed_excess} shows the excess SED of the $\beta$ Leo
system. We find that a 119 K blackbody emission matches the excess
emission for wavelengths longward of 20 \micron. We refer to this 
excess component as dust from the main planetesimal belt, and try to
set constraints using our simple SED model.  We first start to fit the
SED with $a_{min} \sim a_{bl}$, the smallest grains that are bound to
the system against the radiation pressure force, and $a_{max}
\sim$1000 \micron, the largest grain size in our opacity function. In
the case of $\beta$ Leo, $a_{bl}\sim$ 3 \micron.  Note that grains
with sizes larger than 1000 \micron\ contribute insignificantly to the
infrared output due to the combination of the grain properties and the
steep size distribution. In addition to the MIPS, color-corrected
IRAS 60 \micron\ and {\it Herschel} PACS broadband photometry, six
fiducial points (12, 18, 
20, 26.5, 30 and 35 \micron\ as shown in Figure \ref{irs_excess}) from
the IRS spectrum and four points (55.4, 65.6, 75.8 and 86.0 \micron)
from the MIPS-SED data are selected to compute $\chi^2$ in order to
determine the best-fit debris disk model (Table
\ref{photpoints_sed}). We tried both $p$ = 0 and 
$p$ = --1 cases with various combinations of $R_{in}$ and $R_{out}$. The
resultant model emission has a wrong spectral slope in the regions
of 15--25 and 55--95 \micron. We then tried to relax $a_{min}$ to be
larger or smaller than $a_{bl}$ in the fit, and found that the
$a_{min}\sim$5 \micron\ with $p$ = 0 case gives the best $\chi^2$
value ($\chi_{\nu}^2$=0.4). Using the same parameters but with
$a_{min}$= 3 \micron\ gives a $\chi_{\nu}^2$ value of 2.8.  By fixing
these parameters ($a_{min}$ = 5 \micron, $a_{max}$ = 1000 \micron, and
$p$ = 0), we can then derive the best-fit inner and outer radii to be
3$\pm$2 AU and 55$\pm$8 AU, respectively, with a total dust mass of
(1.9$\pm$0.3)$\times$10$^{-4}$ $M_{\earth}$. The SED using these
best-fit parameters is shown in Figure~\ref{sed_excess}.

\begin{deluxetable}{crrl}
\tablewidth{0pt}
\tablecaption{Excess flux densities used in the SED model\label{photpoints_sed}}
\tablehead{
\colhead{$\lambda$} & \colhead{Flux Density} & \colhead{Error} &
\colhead{Source} \\
\colhead{\micron} & \colhead{mJy} & \colhead{mJy} & \colhead{}
}
\startdata 
12.00 &       60.98 &     102.86 &  IRS     \\
18.00 &      256.26 &      53.73 &  IRS     \\
20.00 &      314.69 &      52.34 &  IRS     \\
23.68 &      457.83 &      40.68 &  MIPS    \\
26.50 &      502.92 &      22.57 &  IRS     \\
30.00 &      573.74 &      20.23 &  IRS     \\
35.00 &      705.96 &      47.58 &  IRS     \\
55.36 &      699.79 &      39.58 &  MIPS-SED    \\
60.00 &      598.31 &     195.11 &  {\it IRAS}  \\
65.56 &      718.21 &      66.14 &  MIPS-SED    \\
71.42 &      615.25 &      52.06 &  MIPS    \\
75.76 &      531.73 &      38.21 &  MIPS-SED    \\
85.96 &      502.83 &      75.40 &  MIPS-SED   \\
100.00&      435.58 &      50.02 &  PACS    \\
160.00&      205.27 &      46.00 &  PACS    \\   
\enddata
\end{deluxetable}  

Based on these parameters from the excess SED, we also
construct a model image at 24 \micron\ to compare with the
observations.  We assume the disk mid-plane is aligned with the
stellar equator, inclined by 21.5\arcdeg. The model image was
projected to the inclination angle, and then convolved with model
STinyTim PSFs.  The azimuthally averaged radial profile computed from
the model image is also shown in Figure~\ref{radprof24}, and matches
well with the observed profile at 24 \micron.  This is consistent with
the $\beta$~Leo disk being marginally resolved with a disk radius less
than 80 AU.

\subsection{Inner Warm Belt}

\begin{figure} 
  \figurenum{11}
  \label{null_models_various_rin} 
  \plotone{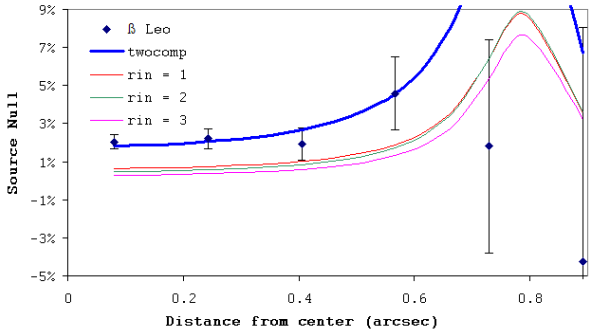}
  \caption{The computed null levels based on the model images for
    $\beta$ Leo. Models based on one main planetesimal belt with various
    inner radii are shown as thin solid lines. Compared to the observed
    null data points (diamonds), these one-component models do not provide
    good fits. The computed null level based on the two-component model
    (main planetesimal belt plus a separate warm belt) is shown as the
    thick solid line, which provides a better fit to the observed data.}
\end{figure}

The next step is to see whether the model we construct for the main
planetesimal belt is consistent with the resolved structure detected
by BLINC. The model excess flux from the main planetesimal belt is
$\sim$57 mJy at 10.5 \micron, which is lower by a factor of 5 compared
to the required level (0.25$\pm$0.05 Jy) to fit the null level derived
from the BLINC data. Reducing the inner radius of the main
planetesimal belt will increase the flux contribution at 10.5 \micron\
(up to $\sim$130 mJy for $R_{in}=$1 AU); however, it also increases
the flux levels in 12--15 \micron\ range, making poor matches with the
observed IRS excess (see Figure \ref{sed_excess}). We also generated
high-resolution model images at 10.5 \micron\ given the various inner
radii and used them as input to compute the BLINC null levels. The
resultant null levels are shown in Figure
\ref{null_models_various_rin}, and they all fail to provide
satisfactory fits inside of 0\farcs6. Therefore, a surface
brightness deficit (gap) in the disk structure is required to explain
both the BLINC and {\it Spitzer} data.

\begin{figure} 
\figurenum{10}
\label{dust_td} 
\plotone{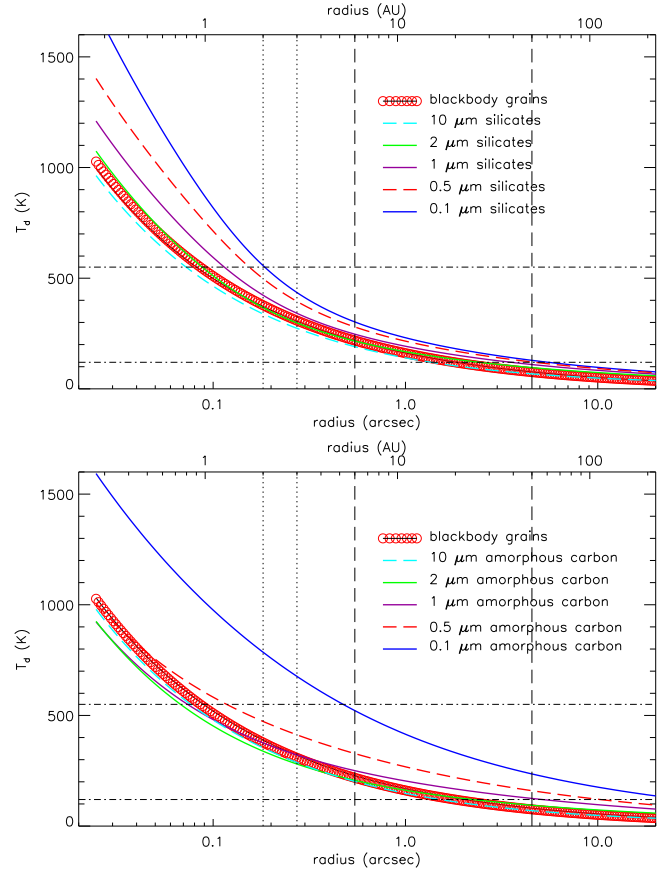}
\caption{Thermal equilibrium dust temperature distribution around
  $\beta$ Leo using grain properties of astronomical silicates (upper
  panel) and amorphous carbon grains (lower panel).}
\end{figure}

We then explore the possibility of an inner warm belt separated from
the main planetesimal belt. Judging from the amount of excess emission
in the range of 5.8--10.5 \micron, the typical dust temperature for
this emission is $\sim$600 K. From the constraints derived from the
BLINC data, we know this warm component is located at 2--3 AU. Figure
\ref{dust_td} shows the thermal equilibrium dust temperature
distribution around $\beta$ Leo. For astronomical silicate grains with
sizes smaller than 1 \micron, the thermal equilibrium temperatures are
generally lower than 500 K at distances of 2--3 AU from $\beta$
Leo. Only sub-micron-size silicates can reach such a high temperature
at a distance of a few AUs. In addition, sub-micron silicate grains have
prominent features near 10 and 20 \micron. Alternatively, sub-micron
carbonaceous grains have equilibrium temperatures of $\sim$600 K, but
produce featureless emission spectrum in the infrared.

\begin{figure}
 \figurenum{12}
  \label{sed_excess_twocomp} 
 \plotone{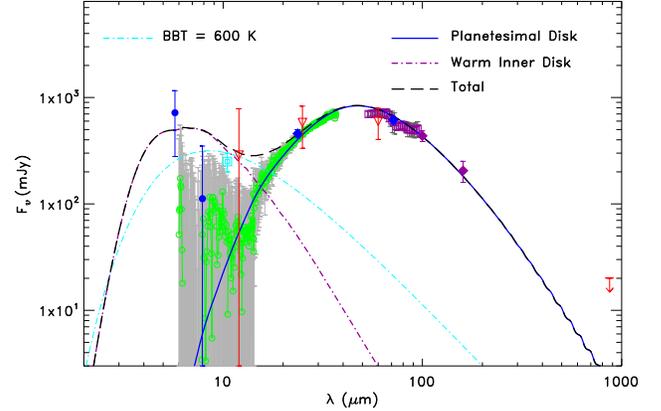}
 \caption{The SED for the final, two-component disk model. Symbols used are the same
as in Figure \ref{sed_excess}}
 \end{figure}

Due to the uncertainty in the exact amount of excess emission for
wavelengths shortward of $\sim$11 \micron, we cannot put real
constraints on grain compositions and sizes. Our goal is to have the
simplest model for the warm belt that can fit the global SED when
combined with the emission from the main belt, and with a null level
computed from the model image that matches the BLINC data. Given the
high dust temperatures in the 2--3 AU belt, we adopted grain
properties for amorphous 
carbon grains (density of 1.85 g cm$^{-3}$, \citealt{zubko96}) with
a radius of 0.5 \micron. The
surface density in this warm belt is assumed to be constant (uniform
distribution at all radii between 2 and 3 AU). With these assumed
parameters, we find that we need very little dust
($M_d\sim$2.6$\times$10$^{-8} M_{\earth}$ or
$f_d\sim7.9\times$10$^{-5}$) to produce the emission shortward of
$\sim$11 \micron. Since this warm belt will contribute some fraction
of the emission longward of $\sim$11 \micron, we have to adjust the
inner radius of the main planetesimal belt to 5 AU so the resultant
combined excess SED fits the observed IRS spectrum within the errors. The
final two-component excess SED is shown in Figure
\ref{sed_excess_twocomp}, and its corresponding null level is shown in
Figure \ref{null_models_various_rin} as well. The final parameters for
the disk are summarized in Table \ref{final_model_param}.

\begin{deluxetable}{ccc}
\tablewidth{0pt}
\tablecaption{Parameters in the two-component model\label{final_model_param}}
\tablehead{ 
\colhead{Parameters} & \colhead{Inner Warm Disk} & \colhead{Planetesimal Disk}  \\
%\colhead{} & \colhead{($warm$)} & \colhead{($pb$)} & \colhead{($halo$)}  
}
\startdata
Adopted Grains   & Carbonaceous & Silicate \\ 
Grain density (g cm$^{-3}$)   &  1.85 & 2.5  \\
Surface density  &  $\sim r^0$  &  $\sim r^0$   \\ 
$R_{in}$ (AU)    &  $\sim$2     &  $\sim$5    \\
$R_{out}$ (AU)   &  $\sim$3     &  $\sim$55    \\
$a_{min}$ (\micron)  &  $\sim$0.5   &  $\sim$5    \\
$a_{max}$ (\micron)  &  $\sim$0.5     & 1000    \\
$M_d$ ($M_{\earth}$)& 2.6$\times$10$^{-8}$ & 2.2$\times$10$^{-4}$ \\ 
$f_d=L_{IR}/L_{\ast}$ & 7.9$\times$10$^{-5}$&2.6$\times$10$^{-5}$  
\enddata 
\end{deluxetable}

\section{Discussion}
\label{discussion}

From the infrared excesses detected from the ground-based
interferometric and {\it Spitzer} data, the debris disk around $\beta$
Leo has two distinct dust components: a main planetesimal belt
distributed from $\sim$5--55 AU that contains second generation dust grains
produced by collisional cascades from large parent bodies residing
in this main belt, and a thin warm belt confined mostly to $\sim$2--3
AU. 
Our simple model (having uniform density distributions in
  both the inner and outer components) suggests that the debris disk around
  $\beta$ Leo has a physical gap ($\sim$3--5 AU) separating the two components.
Since the gap is not near the ice sublimation regions of $\beta$ Leo
($\sim$10 AU), sublimation is unlikely to be the cause for the absence of a measured excess.  
Because the resolved excess emission is only detected at a single wavelength
(N band), we cannot rule out the possibility that the gap is due to a combination of
disk density variations and the resultant temperature structure, which could  cause a surface brightness deficit at 10 $\mu$m, rather than a real physical gap. 
Nevertheless, we believe the gap is most likely due to the presence of an unseen
planetary body. This kind of gap is known to exist in Saturn's ring,
created by the embedded moons directly or by the
orbital resonances of the moons.

It is instructive to compare the detected emission at 10 \micron\, to
the emission from the zodiacal dust disk. Indeed, this comparison is
often used to gauge the level of difficulty in direct imaging of
planets, either via a coronagraph, or an infrared interferometer
\citep{beichman06}.
A model of the solar zodiacal dust provided by \citet{kelsall98}
accurately predicts the emission seen in the DIRBE observations.  For
this comparison, the dust is assumed to be a continuous distribution,
defined by the Kelsall model, with an inner radius due to dust
sublimation ($\sim$0.03 AU for the sun, $\sim$0.15 AU for $\beta$ Leo)
and an outer radius that is defined so as not to affect the amount of
infrared emission (3 AU for the sun, 10 AU for $\beta$ Leo) measured
in the nulled output.

For the BLINC observations a null of 1.7$\pm0.3$\% is fit by a Kelsall
model which is 380$\pm$70 times as dense as the solar zodiacal dust
disk, or 380 zodies.  If the disk were similar in distribution to the
solar zodiacal disk, we would expect KIN to measure a similar null
level (since the disk is well-resolved by both interferometers).
However, the KIN 3 sigma upper limit of 0.9\%, when compared to the
Kelsall model, suggests that $<$ 130 zodies of material is present.
The measurements indicate the current level of knowledge we can obtain
for zodiacal dust around nearby stars for direct imaging, and
highlight the potential danger of using a single "zody" measurement in
characterizing the dust around stars of interest.

The extension of the main dust belt (radius of $\sim$55 AU) around
$\beta$ Leo seems to be small compared to other A-type debris disks at
similar age ($\sim$100 AU scale such as the narrow rings around
HR4796A \citep{schneider09} and Fomalhaut \citep{K05}, or a few 100 AU
scale for the large disks around Vega \citep{su05} and HR8799
\citep{su09}).  The dust mass (summed up to 1000 \micron) is $\sim$10
times less massive than the debris disks around the A-stars Vega and
Fomalhaut, and $\sim$100 times less than the HR 8799 disk, suggesting
that the $\beta$ Leo disk was either born with a low-mass disk or has been through
major dynamical events that have depleted most of the parent bodies.

\citet{akeson09} also reported imaging of $\beta$ Leo
using the Mid-Infrared Echelle Spectrometer (MICHELLE;
\citealt{glasse93}) on the Gemini North telescope, and found the disk
is unresolved at 18.5 \micron. Based on our model image at 18.5
\micron\ (after being convolved with a proper instrumental PSF), the
brightest part of the disk is $\sim$0.55 mJy pixel$^{-1}$ at
$r\sim$0\farcs6 ($\sim$FWHM of a point source) from the star and 0.2
mJy pixel$^{-1}$ at $r\sim$1\farcs2 ($\sim$2 FWHM of a point
source). Given the observational depth ($\sim$0.55 mJy pixel$^{-1}$ on
background), it is not surprising that the disk is not resolved at
18.5 \micron.

Since the longest wavelength that has a sound infrared excess around
$\beta$ Leo is at 160 \micron, our observations are not sensitive to
very cold grains that 
emit mostly in the sub-millimeter and millimeter
wavelengths. Nevertheless, we can set some constraints based on the
{\it Herschel} 160 \micron\ data and the 870
\micron\ and 1.3 mm upper limits from \citet{holmes03}. These upper limits suggest that
$\lesssim$37 K dust can hide in the system without being detected,
which corresponds to a location $>$200 AU for blackbody-like grains. For a
simple uniform-density ring from 200--250 AU consisting of silicate
grains of 100--1000 \micron\, a total dust mass less than
$\sim$3$\times10^{-3} M_{\earth}$ (3-$\sigma$) can exist in this outer
part of the $\beta$ Leo disk. Such a cold ring, if it exists, would have a
clear separation from the main planetesimal belt ($<$80 AU).

\begin{figure}  
\figurenum{13}
\label{radprof70}
\plotone{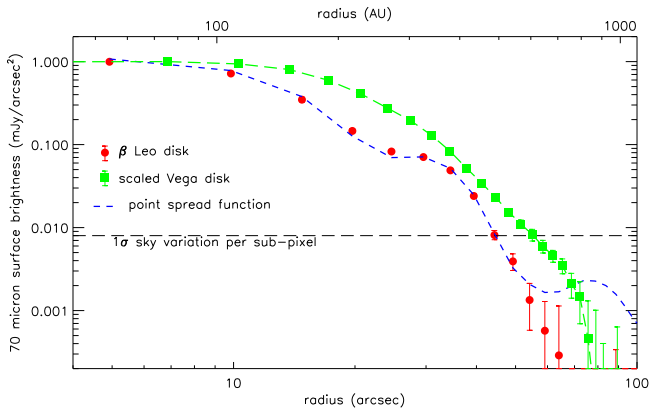}
\caption{The azimuthally averaged radial profile of the $\beta$~Leo
  disk at 70 \micron\ shown in circles compared to the PSF 
  (dashed line). For comparison, we also show the Vega disk profile
  (squares) that has been scaled to match the same distance and the peak surface
  brightness as the $\beta$ Leo disk.} 
\end{figure}

There are 12 A-type debris disks, selected from \citet{S06}, which
have ages between 150 and 400 Myr, spectral types between A0 V and
A3V, and $f_d$ between 10$^{-5}$ and 10$^{-4}$ (for details see
\citealt{S08}). Only the disks around Vega \citep{su05} and $\gamma$
Oph \citep{S08} show large disk extension at 70 \micron\ (radius
$\sim$800 AU for Vega and $\sim$520 AU for $\gamma$ Oph), while the
rest of them have outer disk radii within 200 AU (Su et al.~2010, in prep.).
For a direct comparison, we have rescaled the Vega disk to the same
distance as $\beta$ Leo, matched the peak surface brightnesses in the
disks at 70 \micron, and compared the radial surface brightness
profiles (see Fig.~\ref{radprof70}). If the $\beta$ Leo disk had a halo
similar to the Vega disk, we would have detected it at
$>$10 $\sigma$ levels (at radii of 200--300 AU) around $\beta$
Leo. This suggests that the mechanism that is responsible for creating
such a large halo around Vega does not operate in $\beta$ Leo.

\section{Conclusion}
\label{conclusion}

Using an array of instruments on {\it Spitzer} (imaging, photometry
and spectroscopy) as well as 10 \micron\ nulling interferometry on AU
and sub-AU scales with the MMT and Keck, we have examined the $\beta$
Leonis system and characterized its debris disk. We have found the
system to have at least two distinct components: a warm, narrow ring
located near 2 AU and a broad, cooler ring extending from 5 to 55
AU. We also find the system to lack any significant
belt beyond 80 AU, which is in contrast to many other A-stars with
debris disks. Although not examined in detail here, the
truncation of the outer disk may be indicative of disruption at some
point in its history or the presence of larger planetary bodies,
while the existence of a two-component debris disk may similarly
indicate the presence of planets in the inner parts of the
system.

\acknowledgements

We thank R. Akeson and R. Millan-Gabet for extensive assistance with
the KIN observations. We are greatful to M. Colavita for reduction of
the 2008 Feb. $\beta$ Leo data, J. Bouwman for providing the FEPS IRS 
reduction pipeline and slit off-set correction software, and P. Smith 
for reduction of the MIPS-SED data. We also thank R. Akeson and O. Absil
for discussion of the K-band excess. Some observations reported here
were obtained at the MMT Observatory, a joint facility of the
Smithsonian Institution and the University of Arizona, and by 
the Keck Interferometer, which is funded by the National Aeronautics and
Space Administration as part of its Navigator program.  This work has
made use of services produced by the NASA Exoplanet Science Institute
at the California Institute of Technology. 
%The authors also acknowledge the significant cultural role and reverence that the
%summit of Mauna Kea has always had within the indigenous Hawaiian
%community. 
This work was supported by KI subcontract 1330562 and contract 1255094 from
JPL/Caltech to the University of Arizona.

% \begin{deluxetable}{ccc}
% \tablecaption{Observed and Expected Nulls for {\it o} Leonis \label{tab-omileo_nulls}}
% \tablewidth{0pt}
% \tablehead{
% \colhead{Date} &
% \colhead{Obs. Null (\%)} &
% \colhead{Expect. Null (\%)$^{1}$}
% }
% \startdata
% 16 Feb 2008 & $-0.4 \pm 1.0$ & $2.6 \pm 0.5$\\
% 17 Feb 2008 & $3.6 \pm 0.3$ & $4.3 \pm 0.5$\\
% 14 Apr 2008 & $1.7 \pm 1.0$ & $0.8 \pm 0.5$\\
% \enddata
% \tablecomments{1 - The errors quoted are estimates, incorporating the
%   effects of: 1) the change in interferometer baseline orientation
%   relative to the sky over the time of observation (typically about 1
%   hour); 2) uncertainty in the calculated relative positions of the
%   stellar components; 3) the error in the N-band flux ratio between
%   the stellar components.} 
% \end{deluxetable}

\clearpage

\clearpage

\appendix

\section{Nulling Interferometry with BLINC}

Nulling interferometry
allows light from a central source to be suppressed, while not
affecting surrounding, extended emission. Consequently, fainter
sources surrounding a central source can be more easily observed.

\begin{figure}
\figurenum{A}
\label{fig-TFs}
\epsscale{0.95} 
\plotone{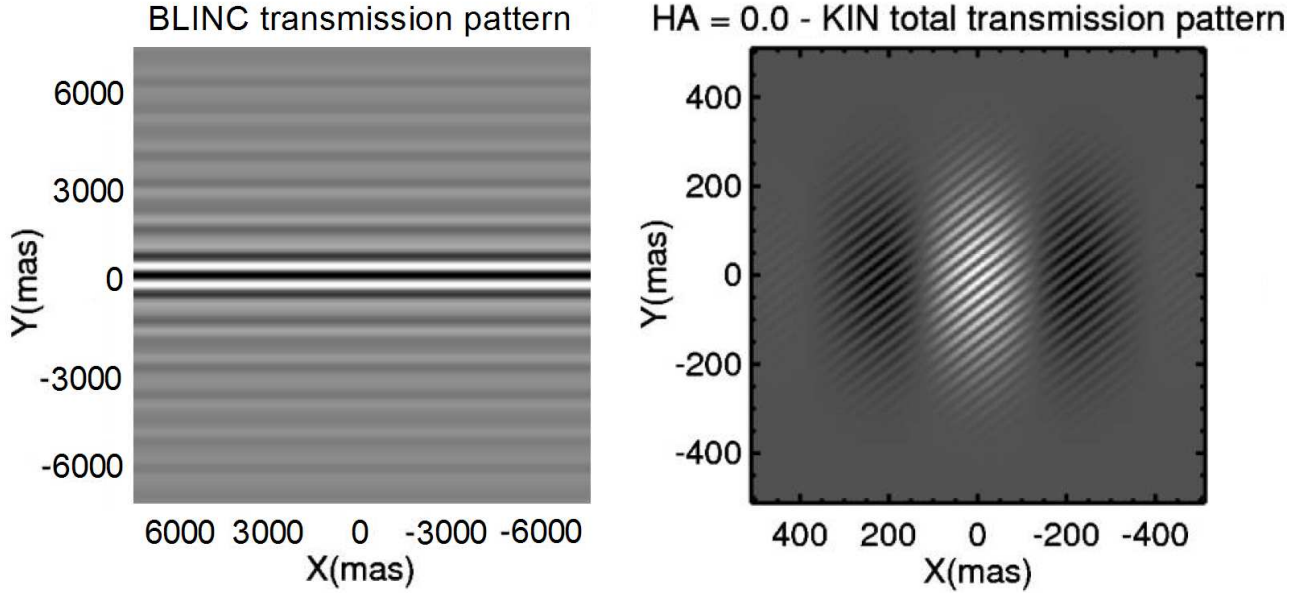}
\caption{(a) The combined N-band transmission pattern for BLINC on the detector.
  (b) The 10 \micron\ transmission pattern for KIN, combining the pupil transmission,
  long baseline fringes and cross-combiner fringes for 0 hour angle and +20 
  degrees declination. From \citet{colavita09}.} 
\end{figure}

Nulling interferometry using BLINC on the MMT operates by splitting
the incoming light into two subapertures. By requiring light from one
subaperture to travel an additional distance of half a wavelength and
then recombining the two lightpaths, a transmission pattern is created
with the form \begin{equation} T(\theta) = \sin{}^2 (\frac{\pi
b\theta}{\lambda}) \end{equation} where $b$ is the interferometer
baseline, $\lambda$ is the wavelength of light and $\theta$ is the
vertical angular distance from the central null. Figure \ref{fig-TFs}a 
shows the transmission pattern created by BLINC for the simplified case 
of uniform contribution from 8 to 13 \micron. Using this pattern, the flux 
from the central star is strongly suppressed while extended structures, 
such as debris disks, are much less affected \citep{H00}. With a baseline 
of 4m (which is the separation of the subapertures used on the MMT), the
first constructive peak occurs at 0\farcs25; however, BLINC is
sensitive to emission outwards of 0\farcs12, where the transmission is
neither destructive nor constructive. This system is ideal for probing
regions containing warm disks around nearby main sequence
stars. Moreover, using the adaptive optics systems on the MMT, we can
obtain stable, diffraction limited images. This allows the pathlength
between beams to remain fixed instead of varying randomly. Such random
variations would require less efficient techniques, such as 'lucky
imaging,' to extract useful data \citep{Hz01}.

To set the appropriate pathlength difference between beams that will
achieve destructive or constructive interference, prior to each
dataset a pair of calibrator frames is taken. A calibrator frame
consists of changing the pathlength to an intermediate distance
(neither constructive nor destructive). For each calibrator pair, one
frame is taken on the shortward side of the optimal pathlength and the
other on the longward side. The observed brightness of the star for
these calibrator frames is directly proportional to the transmissive
efficiency; when BLINC is properly set to destructive interference,
the brightness of the star will be the same for both calibrator frames
($I_1$ = $I_2$). If the pathlength difference is not properly set for
destructive interference, the calibrator frames will be different
brightnesses. To correct BLINC, the optimal pathlength can be
approximated using \begin{equation} Correction = g \frac{I_1 -
I_2}{I_1 + I_2} \end{equation} where $g$ is a constant gain factor which
has been set experimentally using an artificial source. While this
procedure will set BLINC to optimal destructive interference for
moderate offsets, if the initial offset is too large, BLINC may settle
into a side null. These nulls are 15-20\% worse than the central null
while variations from dataset to dataset are much smaller (typically
several percent, see Figure \ref{fig-mmtnulls}). Consequently, data
taken at the incorrect null are easily identified and removed from the
overall dataset.

In order to extract meaningful information, sky subtracted constructive 
and destructive images of both the science star and a calibrator star need 
to be obtained. The "instrumental null" is the null measured by the 
instrument and is the ratio of a destructive image
to a constructive image for a given target. The constructive image allows 
for the normalization of destructive images to a baseline, in this case the 
brightness of the central star. For a monochromatic point source perfectly 
centered in the image, this ratio would be zero, since the target would be 
perfectly nulled; however, in actual observations, stars will only be 
incompletely suppressed, typically yielding a ratio of $\sim$ 3\% , with some 
amount of variation from frame to frame. Thus a
calibrator star is needed to set a baseline with which to compare a
science star (in this case, $\beta$ and o Leo). For such science
stars, the "source null" refers to the instrument null of the science
star subtracted by the instrument null of a calibrator star. If the
science star is unresolved (i.e. a point source), then the source null
will be zero within the error bars. If the star is resolved, then the
source null will be positive. By calculating the source null at
different size scales, spatial information on the object can be
gained.

\section{Interferometric Analysis for {\it o} Leo}

\subsection{KIN} 

We use a simple model, created with the Visibility Modeling Tool (VMT)
provided by the NASA Exoplanet Science Institute.  The tool allows one
to create a hypothetical system, consisting of any number of point
sources, disks and rings that may be either uniform or Gaussian, and
predicts the observational signature if the hypothetical system were
to be observed by KIN.  We used the model to test whether the
variation in null detected can be explained solely by the stellar
components of the binary system.  \citet{H01} determine the necessary
stellar parameters and orbital data necessary to model the system with
two point sources.  We adopt a difference in N-band magnitude between
the primary and secondary of $\delta m_{N} = 1.5$.  Given an orbital
period of 14.5 days for the stellar pair, we can use the date of the
observations to estimate the relative positions of the primary and
secondary to determine a hypothetical observational signature. Values 
calculated for the expected and observed nulls are shown in Table 
\ref{tab-omileo_nulls}. Comparing the two, we can say that the predicted
and actual data are marginally consistent, indicating little evidence for
additional extended emission. 

\begin{deluxetable}{ccc}
\tablecaption{Observed and Expected Nulls for {\it o} Leonis \label{tab-omileo_nulls}}
\tablewidth{0pt}
\tablehead{
\colhead{Date} &
\colhead{Obs. Null (\%)} &
\colhead{Expect. Null (\%)$^{1}$}
}
\startdata
16 Feb 2008 & $-0.4 \pm 1.0$ & $0.87 \pm 0.63$\\
17 Feb 2008 & $3.6 \pm 0.3$ & $1.35 \pm 0.98$\\
14 Apr 2008 & $1.7 \pm 1.0$ & $0.83 \pm 0.60$\\
\enddata
\tablecomments{1 - The errors quoted are estimates, incorporating the
  effects of: 1) the change in interferometer baseline orientation
  relative to the sky over the time of observation (typically about 1
  hour); 2) uncertainty in the calculated relative positions of the
  stellar components; 3) the error in the N-band flux ratio between
  the stellar components.} 
\end{deluxetable}

Given the behavior of the null with respect to time as compared to the
detailed orbital parameters determined by \citet{H01}, we can conclude
that the non-zero nulls detected by our KIN observations are likely due to
the stellar components being resolved, with no evidence for excess
emission from warm dust.  A conservative 3 $\sigma$ upper limit for 10
$\mu$m emission from hot dust in the system can be taken to be 3\%.

\subsection{BLINC} 

Although there was no significant excess emission around {\it o} Leo,
we can set constraints on what dust might be present there. {\it o}
Leo is a spectroscopic binary located at a distance of 41.5 pc with a
separation of 0.21 AU (making the semi-major axis 5 mas). The orbit has
a 57.6 degree inclination from face on \citep{H01}, and we will make
the assumption that any circumbinary debris disk will have a similar
inclination to us.

If we assume blackbody grains and the above parameters for the system,
we can calculate the radius where we would expect to find 300K
grains. This is the temperature at which the blackbody emission peaks
around 10 \micron. The physical parameters for the host stars are
derived to be 5.9\rsun\ and 6000 K for the first star and 2.2\rsun\ and
7600 K for the second star \citep{H01}. For spherical blackbody grains,
\begin{equation} T_g = (\frac{T_1^4 R_1^2 + T_2^4 R_2^2}{2r^2})^{.25}
\end{equation} So $T_{g}$ = 300 K when $r$ = 9 AU, which for {\it o} Leo 
is at an angular separation of 0\farcs22 along the semi-major
axis. Assuming the disk is oriented vertically on the detector (and
thus perpendicular to the interference pattern), this would put such a
ring close to the constructive peak, making BLINC especially sensitive
to any dust located in that region.

\begin{figure}
\figurenum{B}
\label{fig-oRingMod}
\epsscale{0.95} 
\plotone{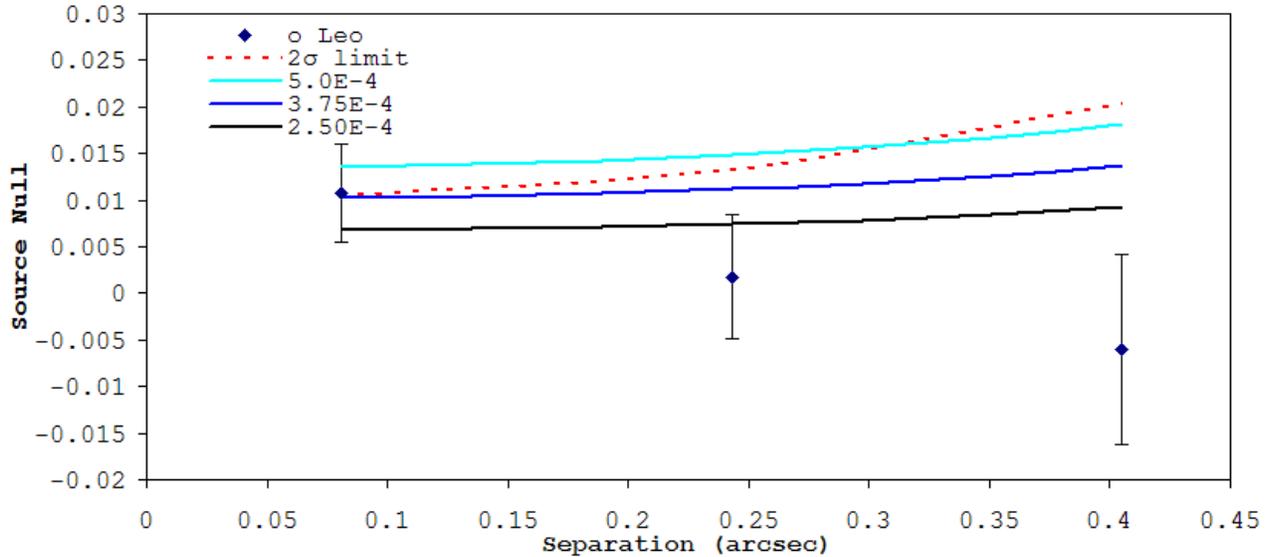}
\caption{Ring models (8--10 AU) of varying optical depth for {\it o}
  Leo. The semi-major axis is aligned perpendicular to the
  interference pattern. The dotted line represents the 2 $\sigma$
  detection limit based on our observations. Also shown is the source
  null for {\it o} Leo, originally plotted in Figure 2b.}  
\end{figure}

Figure \ref{fig-oRingMod} shows the source null of several ring models
plotted along with the {\it o} Leo datapoints. The models assume the
orientation of the disk is perpendicular to the interference pattern
of the interferometer. The optical depth of each model is listed in
the legend, and the rings extend from 8 to 10 AU, directly through the
300 K region around the star. Plotted as a dotted line is the 2 $\sigma$
limit (based on the error bars of each datapoint) above which we would
have a marginal detection. Thus, if there were a dust ring equivalent
to the 3.75\sci{-4} model in Figure \ref{fig-oRingMod}, we would not
expect to have a positive detection, whereas we would expect to have
gotten a marginal detection if the dust ring had an optical depth of
5\sci{-4}. Since models that are brighter than $\sim$3.75\sci{-4}
begin to lie above the 2 $\sigma$ threshold, this model marks the upper
limit we can place on dust in such a ring, which corresponds to a flux
of 0.17 Jy at 12 \micron.

If, on the other hand, the disk were oriented parallel to the
interference pattern, a brighter disk could be present without being
detected. For this sub-optimal orientation, an upper limit of 0.31 Jy
can be placed on the disk.

\subsection{Comparison to previous measurements}

{\it o} Leo has previously been identified as having a very hot excess
by \citet{T07} using MIPS on {\it Spitzer}. \citet{T07} reported a 24
and 70 $\mu$m excess ratios for {\it o} Leo of 1.23 and 1.30
respectively, suggesting the presence of circumbinary dust.  They
derived a dust temperature of 815 K with a minimum radius of 0.85 AU
(0\farcs02). Although too concentrated to be detectable by BLINC, such
a disk would be odds with the lack of detection by KIN. However,
Trilling et al's analysis appears to be based on saturated 2MASS
measurements for {\it o} Leo, from which they derive a K-band magnitude
of 2.58 $\pm$ 0.15. If instead we use Johnson K-band photometry \citep{johnson66}
converted to a 2MASS K$_S$ magnitude, we find that {\it o} Leo is 2.39
$\pm$ $<$0.07, 0.19 magnitudes brighter than calculated by Trilling et
al. Since the 2MASS data is saturated for {\it o} Leo, we believe this
Johnson K-magnitude to be a better measurement. Using this magnitude, our model fitting predicts flux densities of 816 and
89.5 mJy (or ratios of 1.00 and 1.06 to the stellar photosphere)
respectively at 24 and 70 \micron. The MIPS measurements 
reported by \citet{T07} then indicate no excess at 24 $\mu$m above the
$\sim$ 6\% (2-$\sigma$) level (and above 15\% at 70 $\mu$m). The new ratios indicate that
the system harbors little, if any, dust. The lack of excess
detections at longer wavelengths is then consistent with the null
detections by BLINC and KIN at 10 \micron.

\end{document}